\documentclass[12pt]{article}
\usepackage{graphicx,lscape}

\def\beq{\begin{equation}}   
\def\eeq{\end{equation}}
\def\lsim{\mathrel{\rlap{\lower3pt\hbox{\hskip0pt$\sim$}}
    \raise1pt\hbox{$<$}}}         
\def\gsim{\mathrel{\rlap{\lower4pt\hbox{\hskip1pt$\sim$}}
    \raise1pt\hbox{$>$}}}         

\begin{document}

\begin{titlepage}

\begin{flushright}
TPI-MINN-02/6\\
UMN-TH-2046/02\\

\end{flushright}

\vspace{0.3cm}

\begin{center}
\baselineskip25pt

{\Large\bf Duality and Self-Duality (Energy Reflection
Symmetry) of  Quasi-Exactly Solvable Periodic Potentials}

\vspace{0.3cm}

{ \bf  Gerald V. Dunne}

{\em Department of Physics, University of Connecticut\\ 
2152 Hillside Road, Storrs, CT 06269}

\vspace{0.1cm}

{ \bf M.~Shifman}

{\em Theoretical Physics Institute, University of Minnesota, \\
116 Church St. S.E., Minneapolis, MN 55455}

\end{center}

\vspace{1cm}
\begin{center}
{\large\bf Abstract} \vspace*{.25cm}
\end{center}

A class of spectral problems with a hidden Lie-algebraic structure is considered.
We define a duality transformation which maps the spectrum of one quasi-exactly
solvable (QES) periodic potential to that of another QES periodic potential. The
self-dual point of this transformation corresponds to the energy-reflection
symmetry found previously for certain QES systems. The duality transformation
interchanges bands at the bottom (top) of the spectrum of one
potential with gaps at the top (bottom) of the spectrum of the other, dual,
potential. Thus, the duality transformation provides an exact mapping between the
weak coupling (perturbative) and semiclassical  (nonperturbative) sectors.

\end{titlepage}

\newpage

\section{Introduction}
\label{intro}

Quasi-exactly solvable (QES) systems are those for which some finite portion of
the energy spectrum can be found exactly using algebraic means
\cite{turbiner,ush,st1,olver,itep}. A positive integer parameter $J$
characterizes the `size' of this exact portion of the spectrum. This integer $J$
has both an algebraic significance, related to the dimension of a representation, 
and a geometrical significance, the genus of a Riemann surface associated with
the spectrum. In \cite{bdm}, the large $J$ limit was identified as a
semiclassical limit useful for studying the top of the quasi-exact spectrum. It
was found that remarkable factorizations reduce the semiclassical calculation to
simple integrals, leading to a straightforward asymptotic series representation
for the highest QES energy eigenvalue. The notion of energy-reflection (ER)
symmetry was introduced and analyzed in \cite{st2}: for certain QES systems the
QES portion of the spectrum is symmetric under the energy reflection $E\to - E$.
This means that for a system with ER
symmetry, there is a precise connection between the top of the QES spectrum and
the bottom of the spectrum. Coupled with the semiclassical large $J$ limit,  the ER
symmetry relates semiclassical (nonperturbative) methods with perturbative
methods \cite{st2}. In this present paper, we show that for a class of periodic QES
potentials the ER symmetry is in fact the fixed point (self-dual point) of a more
general duality transformation. The duality transformation we consider is analogous
to, but importantly different from, a similar transformation suggested in
Ref.~\cite{krajewska} relating different QES potentials. 

For
periodic potentials, usually, the Lie-algebraic QES sector consists of 
band boundaries which correspond  either to periodic or antiperiodic
boundary conditions~\footnote{The simplest periodic QES problems were
discussed in the literature long ago
\cite{razavy}; in particular, the Lam\'{e} system 
was the subject of investigation in Refs. \cite{turbiner2,Gturbiner2}. }. 
Thus, only one boundary of each band is algebraic. However,
for the elliptic periodic potentials to be considered below,
both boundaries are algebraically calculable.
For given $J$, there are $2J+1$ energy eigenvalues in the algebraic
sector; there are
 $J+1$ allowed
bands,  with the highest band being open (it stretches up to $E=\infty$).
The $2J+1$ algebraic energy levels
give {\em all}  band edges, so that in the problem at hand {\em all} band
edges   are algebraically calculable. This fact is well-known in the
literature. In this paper the main emphasis is on physical aspects which
have not been investigated so far. When the parameter
$J$ becomes large, the theory becomes weakly coupled at the bottom of the
spectrum and semiclassical at the top. We develop this idea --- using $1/J$ as an
expansion parameter --- in the application to the elliptic periodic potentials. 
We show that the
duality between weak coupling and semiclassical expansions
 applies not just to the
asymptotic series for the locations of the bands and gaps, but also to the
exponentially small widths of bands and gaps. We observe that
the $2J+1$-level algebraic sector splits into four completely
disconnected  algebraic subsectors. Therefore, instead of diagonalizing a
$(2J+1)\times (2J+1)$ matrix, it is sufficient to diagonalize
four matrices which are roughly four times smaller in their linear
sizes. In addition to the eigenvalues,
we determine, purely algebraically, the eigenfunctions
corresponding to the band edges. For even $J$ there are
$J+1$ periodic eigenfunctions and $J$ antiperiodic ones. For odd $J$
there are $J$ periodic eigenfunctions and $J+1$ antiperiodic ones.

In Sect. \ref{introduality} we define our duality and self-duality transformation,
and review some basic facts about the periodic Lam\'e equation. The duality
transformation has both an algebraic and geometric interpretation. 
We show how the duality transformation extends 
the algebraic ER
symmetry formalism introduced in
\cite{st2}.
Section~\ref{QES} 
is devoted to a detailed consideration of four
distinct $sl(2)$-based algebraizations relevant to the algebraic determination
of the band edges and the corresponding wave functions.
We explain why the cases of even and odd $J$ must be treated
 separately and how periodic and antiperiodic wave functions
emerge. The periodicity {\em versus} antiperiodicity
is due to properties of the quasiphases in these two cases.
In Sect.~\ref{PE} we study several explicit examples in detail.  
Section~\ref{nu=1/2} is devoted to an analysis of the self-dual case. In
Sect.~\ref{nonpert} we describe the duality relation between perturbative and
nonperturbative approximation techniques, as applied to the computation of both
the locations and widths of bands and gaps. Finally, Sect.~\ref{conclu} contains a
summary and some comments regarding further generalizations.

\section{Duality and self-duality in Lam\'e model}
\label{introduality}

\begin{figure}[htb]
\begin{center}
\includegraphics[width=15cm]{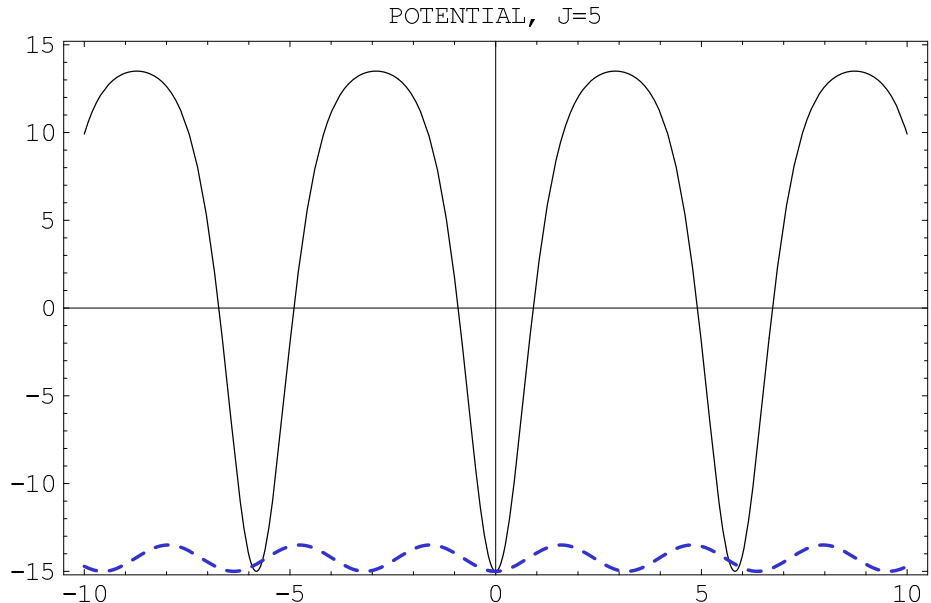}
\end{center}
\caption{
The potential energy in the Schr\"odinger equation (\protect{\ref{lame}})
as a function of $\phi$.  The solid curve corresponds to elliptic parameter $\nu
=0.95$, for which the period is $2K(0.95)\approx 5.82$. The dashed curve
corresponds  to $\nu = 0.05$, for which the period is $2K(0.05)\approx 3.18$. In
each case, $J=5$. Note how different the two potentials are; and yet, their
spectra are related by the duality transformation (\protect{\ref{duality}}).}
\label{ppf6}
\end{figure}

In this section we introduce the notions of duality and self-duality for the 
following one-dimensional quasi-exactly solvable (QES) Lam\'e equation:
\beq
\left\{ -\frac{d^2}{d\phi^2} +J(J+1)\,\nu\,{\rm sn}^2(\phi|\nu )- 
\frac{1}{2}J(J+1)\right\}\Psi (\phi) = E\, \Psi (\phi)\,.
\label{lame}
\eeq
Here ${\rm sn}(\phi|\nu )$ is the doubly-periodic Jacobi elliptic  function
\cite{ww,abramowitz}, the coordinate $\phi\in R^1$, and $E$ denotes
the energy eigenvalue. The real elliptic parameter $\nu$ lies in the range
$0\leq\nu\leq 1$.  The potential in (\ref{lame}) has period $2K(\nu)$, where 
\begin{eqnarray}
K(\nu)=\int_0^{\pi/2} \frac{d\theta}{\sqrt{1-\nu \sin^2\theta}}
\label{qperiod}
\end{eqnarray}
is the elliptic quarter period.
Note that the parameter
$\nu$ controls the period of the potential, as well as its strength. As
$\nu\to 1$, the period $2K(\nu)$ diverges logarithmically, $2K(\nu)\sim \ln\,
(\frac{16}{1-\nu})$, while as $\nu\to 0$, the period tends to a nonzero
constant: $2K(\nu)\to \pi$. In the Lam\'e equation (\ref{lame}), the
parameter $J$ is a positive integer (for non-integer $J$, the problem is not
QES). This parameter $J$ controls the depth of the wells of the potential;
 the significance of the constant subtraction $-\frac{1}{2}J(J+1)$ will
become clear below. As an illustration, we have plotted in Fig. \ref{ppf6}
the potential energy in the Schr\"odinger equation (\ref{lame}) for two
values of $\nu$, namely, $\nu = 0.95$ and $\nu =0.05$.

It is a classic result that the Lam\'e equation (\ref{lame}) has bounded solutions
$\Psi(\phi)$ with an energy spectrum consisting of exactly
$J$ bands, plus a continuum band \cite{ww}. It is the simplest example of a
``finite-gap" \cite{mckean,belokolos} model \footnote{Much of our discussion
generalizes to other finite gap potentials, but we concentrate here on the
Lam\'e system for the sake of pedagogical definiteness.}, there being just a
finite number, $J$, of ``gaps", or ``excluded bands" in the spectrum. This should
be contrasted with the fact that a generic periodic potential has an infinite
sequence of gaps in its spectrum \cite{magnus}. We label the band edge energies by
$E_l$, with $l=1,2,\dots, (2J+1)$. Thus, the energy regions, $E_{2l-1}\leq E\leq
E_{2l}$, and $E\geq E_{2l+1}$, are the allowed bands (``bands" for short), while
the regions,
$E_{2l}< E < E_{2l+1}$, and $E< E_1$, are the exclusion bands (``gaps" for short).
The wave functions
$\Psi(\phi)$ at the band edges are either periodic or antiperiodic functions,
with period
$2K(\nu)$, while within the bands the wave functions are quasi-periodic Bloch
functions, which can be expressed in terms of theta functions \cite{ww}.

\begin{figure}[htb]
\includegraphics[width=15cm]{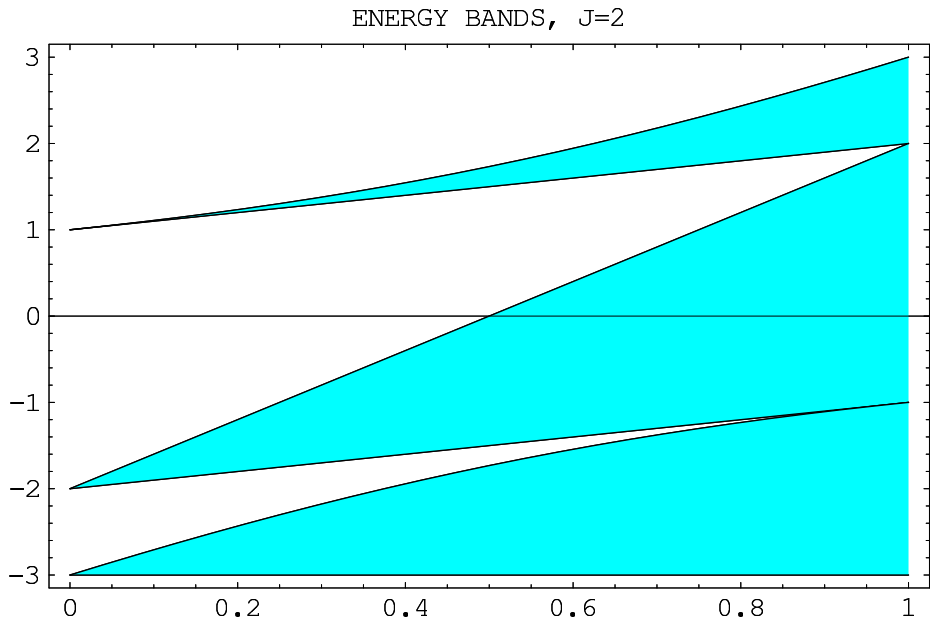}
\caption{
The energy bands for the Lam\'e system (\ref{lame}), as a
function of $\nu$. The
plot given is for $J=2$. The first (the lowest)   allowed band (unshaded)
is of the p-ap type, while the second one of the ap-p type, see text.}
\label{ppf3}
\end{figure}

Another important classic result \cite{iachello,ward,kusnezov} concerning
the Lam\'e model (\ref{lame}) is that the band edge energies $E_l$, for
$l=1,\dots, 2J+1$, are simply the eigenvalues of the $(2J+1)\times(2J+1)$
matrix
\begin{eqnarray}
H=J_x^2+\nu J_y^2 -\frac{1}{2} J(J+1)\, {\bf I}
\label{ham}
\end{eqnarray}
where $J_x$ and $J_y$ are $su(2)$ generators in a spin $J$ representation
and ${\bf I}$ is the unit matrix.
Thus the Lam\'e band edge spectrum is {\it algebraic}, requiring only the
finding of the eigenvalues of the finite dimensional matrix $H$ in (\ref{ham}).
When $\nu= 0\,\, {\rm or} \,\, 1$, it is straightforward to write down the
eigenvalues of $H$. But for $0< \nu <1$, the matrix $J_x^2+\nu J_y^2$ cannot
be simply related to the $su(2)$ Casimir, $\vec{J}^2$, and so the
eigenvalues of $H$ are, in fact, nontrivial functions of $\nu$. For
example, for $J=1$ and $J=2$, the eigenvalues of $H$ are:
\begin{eqnarray}
J=1\quad :\quad E_1&=&-1+\nu \, ,\nonumber\\
E_2&=& 0\,,\nonumber\\ 
E_3&=& \nu\,;\nonumber \\[2mm]
J=2\quad :\quad 
E_1&=&-1 + 2\,\nu - 2\,{\sqrt{1 - \nu +\nu^2}}\,,\nonumber \\
E_2&=& -2 + \nu\,,\nonumber \\
E_3&=& -2 + 4\,\nu  \,,\nonumber \\
E_4&=&1 + \nu\,, \nonumber \\
E_5&=&-1 + 2\,\nu + 2\,{\sqrt{1 - \nu + \nu^2}}\,.
\label{egs}
\end{eqnarray}
The band (gap) structure in the second case, $J=2$, is presented in Fig.
\ref{ppf3}.

The shaded areas on this plot (and other similar plots in Figs. \ref{ppf4},
\ref{ppf1} and \ref{ppf2}) are
the forbidden bands (gaps), while the unshaded areas are the allowed bands.
The allowed bands can be of two types: p-ap and ap-p. In the allowed bands
of the first type the lower boundary of the band is determined by a
periodic wave function, while the upper boundary by an anti-periodic one.
In the allowed bands of the ap-p type, the band edge structure is
reversed:  the lower boundary corresponds to an
 anti-periodic  wave function. The last allowed band has no upper boundary ---
it stretches up to infinitely high energies.

In \cite{dr}, the algebraic form (\ref{ham})  of the Lam\'e system was exploited to
provide a precise analytic test of the instanton approximation, as is
discussed further in Sect.~\ref{nonpert}.

In this paper we study a special duality of the spectrum of the Lam\'e system
(\ref{lame}), which can be stated succinctly as:
\begin{eqnarray}
E[\nu]=-E[1-\nu]
\label{duality}
\end{eqnarray}
where $E[\nu]$ denotes the spectrum for the potential with elliptic parameter
$\nu$.  That is, the spectrum of the Lam\'e system (\ref{lame}), with elliptic
parameter $\nu$, is the energy reflection of the spectrum of the Lam\'e
system with the dual elliptic parameter $1-\nu$. In particular, for the band
edge energies, $E_l$, which are the eigenvalues of the finite dimensional
matrix $H$ in (\ref{ham}), this means that (for $l=1,2,\dots, 2J+1$)
\begin{eqnarray}
E_l[\nu]=-E_{2J+2-l}\,[1-\nu]
\label{hduality}
\end{eqnarray}
This duality can be seen directly in the eigenvalues of the $J=1$ and $J=2$
examples in (\ref{egs}). The proof for the band edge energies is a trivial
consequence of the algebraic realization (\ref{ham}), since
\begin{eqnarray}
H[\nu]&\equiv &J_x^2+\nu J_y^2 -\frac{1}{2} J(J+1)\, {\bf I}\nonumber\\
&=&  -\left[
\left(J_z^2+(1-\nu)J_y^2\right)-\frac{1}{2} J(J+1)\,{\bf I}\right]\,.
\label{proof}
\end{eqnarray}
Noting that $[J_z^2+(1-\nu)J_y^2]$ has the same eigenvalues as
$[J_x^2+(1-\nu)J_y^2]$, the duality result (\ref{hduality}) follows. It is
instructive to see this duality in graphical form, by looking at Figs.~\ref{ppf3},
\ref{ppf4}, \ref{ppf1} and \ref{ppf2},
 which show the band spectra, as a function of $\nu$, for various
different values of $J$. In each case, the transformation $\nu\to 1-\nu$,
together with the energy reflection $E\to -E$, interchanges the shaded
regions (the gaps) with the unshaded regions (the bands). 

The fixed point, $\nu=\frac{1}{2}$, is the ``self-dual" point, where the
system maps onto itself, and the energy spectrum has an energy reflection
(ER) symmetry, as was studied in \cite{st2}. In Sects. 3 and 4 we discuss
in detail how this Lam\'e model fits into the energy reflection symmetry
classification of Shifman and Turbiner.

\begin{figure}[htb]
\begin{center}
\includegraphics[width=13cm]{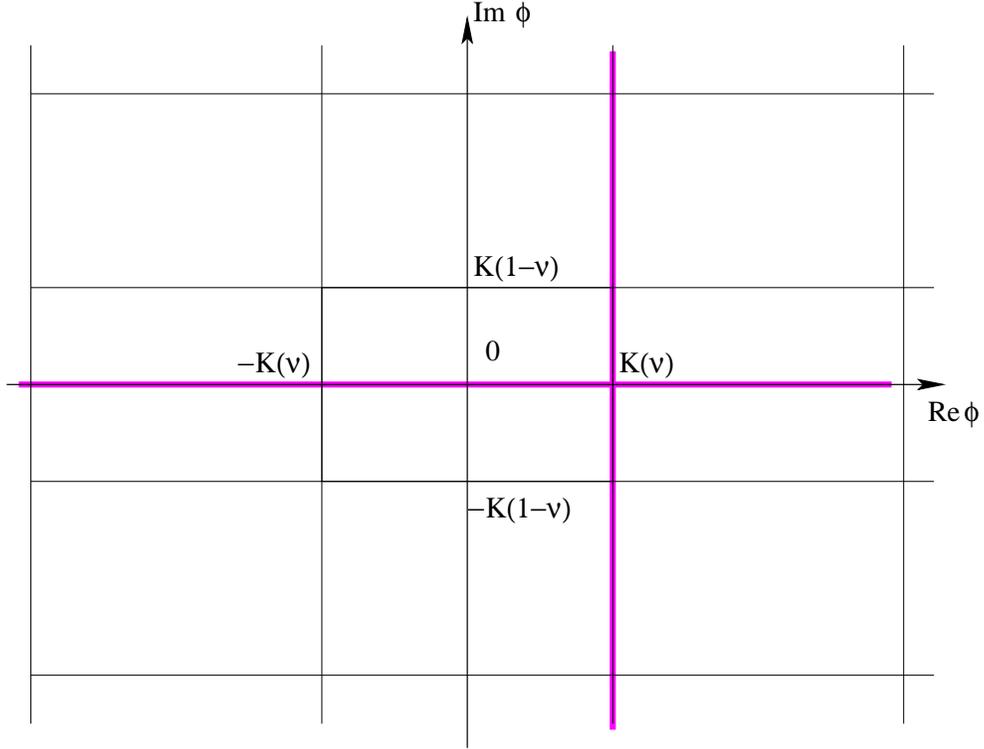}
\end{center}
\caption{
The period parallelogram for
${\rm sn}^2(\phi|\nu )$ is the rectangle with the sides $2K(\nu )$ and
$2K' (\nu )\equiv 2K(1-\nu )$. The original Lam\'e equation (\ref{lame}) is defined
for real $\phi$, on the horizontal thick line. The dual Lam\'e equation is defined
for $\phi$ on the vertical thick line. The parametric coordinate rotation is
(\ref{rotation}), and leads to the dual Lam\'e equation (\ref{lameprime}).}
\label{ppf9}
\end{figure}

In fact, the duality relation (\ref{duality}) applies to the entire
spectrum, not just the band edges (\ref{hduality}). The easiest way to see this is
to realize that the duality relation is a simple consequence of Jacobi's imaginary
transformation, applied to the Lam\'e equation (\ref{lame}). The original Lam\'e
system (\ref{lame}) was defined for real coordinate $\phi$, but we can
extend $\phi$ into the complex plane. The potential is doubly
periodic, with real period $2K$ and imaginary period $2K^\prime\equiv
2K(1-\nu)$, as is shown in Fig.~\ref{ppf9}. Consider rotating the coordinate
off the real axis by the transformation
\begin{eqnarray}
\phi^\prime=i\left(\phi-K-i\, K^\prime\right)\,,
\label{rotation}
\end{eqnarray}
so that real $\phi^\prime$ can be considered as a parametric coordinate for
$\phi$ running along a vertical line through the point $K$ (the vertical thick line
in Fig.~\ref{ppf9}). Note that along this vertical line the potential ${\rm
sn}^2(\phi|\nu)$ is again real. Moreover, using the properties of the Jacobi elliptic
functions, 
\cite{ww,abramowitz}, we get
\begin{eqnarray}
\nu\, {\rm sn}^2(K+iK^\prime - i\phi^\prime |\nu)=1-(1-\nu)\, {\rm
sn}^2(\phi^\prime |1-\nu)\,.
\label{jacobi}
\end{eqnarray}
The Lam\'{e} equation on the real axis maps to another, dual, Lam\'e equation on the
vertical thick line (see Fig. \ref{ppf9}). In other words, 
 the change of variables (\ref{rotation}) transforms the Lam\'e
equation (\ref{lame}) into:
\beq
\left\{ -\frac{d^2}{d\phi^{\prime 2}} +J(J+1)\,(1-\nu)\,
{\rm sn}^2(\phi^\prime|1-\nu)- 
\frac{1}{2}J(J+1)\right\}\Psi (\phi^\prime) =- E\, \Psi (\phi^\prime)\,,
\label{lameprime}
\eeq
So solutions of the Lam\'e equation (\ref{lame}) are mapped to solutions of
the dual equation (\ref{lameprime}), $\nu\to 1-\nu$, and with a sign
reflected energy eigenvalue: $E\to -E$. Returning to the complex $\phi$
plane diagram in Fig.~\ref{ppf9}, the transformation (\ref{rotation}) clearly
interchanges the real and imaginary periods. In the self-dual case,
$\nu=\frac{1}{2}$, these two periods are equal, so the periodic lattice
shown in Fig.~\ref{ppf9} is a square lattice. This is the so-called {\it
lemniscate} case \footnote{Note that $K(1/2) =\tau/\sqrt{2}$ where $\tau$
is the parameter defined in Eq. (\ref{tauperiod}).}.

It is important to distinguish our duality transformation from another duality
transformation considered in \cite{krajewska}, for which the duality
transformation was $\phi^\prime=i\,\phi$, which maps the real axis onto the
imaginary axis. The Lam\'e potential is also real along the imaginary axis.
However, along the imaginary axis there is a singularity at $\phi=i\, K^\prime$. 
The transformation $\phi^\prime=i\,\phi$ maps the original Lam\'e potential to a very
different potential; so there is no notion of ``self-duality" for this
transformation.

To see why bands and gaps are interchanged under our duality transformation
(\ref{rotation}), we recall that the two independent solutions of the original
Lam\'e equation (\ref{lame}) can be written  as products of theta functions
\begin{eqnarray}
\Psi_\pm
(\phi)=
\prod_{j=1}^J\left[\frac{\vartheta_1(\phi\pm\alpha_j)}{\vartheta_4(\phi)}\,
\exp\left(\mp \phi\, Z(\alpha_j)\right)\right]
\label{thetas}
\end{eqnarray}
where the parameters $\alpha_j$, for $j=1,\dots ,J$, satisfy a complicated
set of $J$ nonlinear constraints together with the energy $E$, and
$Z(\alpha)$ is the Jacobi zeta function \cite{ww}. Under the change of
variables (\ref{rotation}) these theta functions map into the same theta
functions, but with dual elliptic parameter. However, they map from bounded
to unbounded solutions (and vice versa), because of the ``i" factor
appearing in (\ref{rotation}). Thus, the bands and gaps become
interchanged. This interpretation of the duality transformation in terms of
the Jacobi imaginary transformation will be important in Sect.
\ref{nonpert} when we discuss WKB techniques.

Away from the self-dual point, $\nu=\frac{1}{2}$, the duality transformation
relates the spectrum of one elliptic potential with the spectrum of a {\it
different} elliptic potential. For example, the two potentials plotted in 
Fig.~\ref{ppf6}, for $\nu=0.95$ and $\nu=0.05$, have energy spectra that are
inversions $E\to -E$ of one another. It is striking how different these two potentials
are, even though their spectra are related by this simple energy-reflection
symmetry. In the semiclassical large $J$ limit, the duality symmetry (\ref{duality})
connects properties of low-lying bands of one potential with high-lying
gaps of the dual potential. At large $J$, the barriers between neighboring
potential wells become high, so that tunneling effects are suppressed.
Thus, low-lying {\it bands} are exponentially narrow, so that we can study
their location and their width. Duality relates these results to the
location  and width of high-lying {\it gaps} for the dual potential. 

There is another interesting semiclassical limit that can be studied for the
Lam\'e models. Since the elliptic parameter $\nu$ controls the period, $2K(\nu)$,
of the potential, the duality transformation relates a system with $\nu\to 1$,
for which the individual wells are far separated (recall that $K(\nu)\to\infty$
as $\nu\to 1$), with a dual system having elliptic parameter $\nu^\prime=1-\nu\to
0$, for which the wells are shallow and close together. In the first case
tunneling is suppressed since the neighboring wells are far apart, so that
semiclassical techniques are appropriate. On the other hand, in the dual system
the potential becomes weak so that perturbative techniques can be used. The
duality relation provides a direct mapping between these different approximate
methods, as is studied in Section \ref{nonpert}.

\section{ Algebraic approach to QES spectral problem with elliptic potentials --
generalities}
\label{QES}

In this section we show how the duality and self-duality properties are
characterized using the algebraic language of QES systems. In accordance
with the general strategy of generating $sl(2)$-based QES   spectral
problems \cite{st2}, we introduce a new variable
$\eta (\phi ) $ satisfying the condition
\beq
\left(\frac{d\,\eta}{d\,\phi}
\right)^2 = 4\, (1- \eta)\,\eta\, \left[(1-\nu )+\nu\,\eta
\right]\,.
\label{twos}
\eeq
The solution of this condition appropriate for our purposes
is
\beq
\eta = 1 - {\rm sn}^2(\phi|\nu )\,.
\label{threes}
\eeq
For real $\phi$ the variable $\eta$ varies between 0 and 1.
As we will see shortly, of relevance are functions $\Psi (\eta )$
which are constructed of polynomials of $\eta$ and may have
square root singularities of the type $\sqrt{\eta}$ or $\sqrt{1-\eta}$;
these functions are either periodic or antiperiodic in $\phi$, and can be
used to construct band-edge wave functions.

In terms of the variable $\eta$, three $sl(2)$ generators
have a differential realization
\begin{eqnarray}
T^+ = 2\,j\,\eta -\eta^2\,d_\eta\,,\qquad
T^0= - j\,\eta + \eta\, d_\eta\,,\qquad
T^-= d_\eta\,
\label{four}
\end{eqnarray}
where 
\begin{eqnarray}
[T^+,T^-]=2T^0\,,\qquad
[T^+,T^0]= -T^+\,,\qquad
[T^-,T^0]= T^-\,,
\label{five}
\end{eqnarray}
and $d_\eta \equiv \frac{d}{d\,\eta}$.
If $j$ is semi-integer, the algebra (\ref{five}) has a finite-dimensional
representation, of dimension $2j+1$,
\beq
R_j=\{ \eta^0\,,\,\,  \eta^1\,,\,\, ...,\eta^{2j}\}\,.
\eeq
The precise relation between the parameter $j$ appearing in (\ref{four})
and the QES parameter $J$ will be discussed in detail in the following.

 It
is convenient to collect in one place the inverse relations, connecting
$\eta^p\, d_\eta^{\, q}$ to the generators (\ref{four}). These relations are
\begin{eqnarray}
\eta^3\, d_\eta^2&=& -T^+T^0 -(3j-1)T^+ +2j(2j-1)\, \eta\,,\nonumber\\
\eta^2\, d_\eta^2&=& -T^+T^- + 2j \, T^0 +2j^2\,,\nonumber\\
\eta\, d_\eta^2&=& T^0T^- + j \, T^- \,, 
\end{eqnarray}
and
\begin{eqnarray}
\eta^2\, d_\eta &=&-T^+  +2j\, \eta \,,\nonumber\\
\eta \, d_\eta &=&   T^0 + j \,,\nonumber\\
  d_\eta &=&   T^- \,.
\end{eqnarray}
The matrix realization of the algebra (\ref{five}) for  semi-integer $j$,
which  will be useful for our purposes,
is well-known:
\begin{eqnarray}
&& T^+ =\left[
\begin{array}{cccccc}
0 & 1 & 0 & ... & 0 & 0\\
0 & 0 & 2 & ... & 0 & 0\\
0 & 0& 0 & ... & 0 & 0\\
... & ...& ... & ... & ... & ...\\
0 & 0 & 0 & ... & 0 & 2j\\
0 & 0 & 0 & ... & 0 & 0
\end{array}
\right],
\quad
T^- =\left[
\begin{array}{cccccc}
0 & 0 & 0 & ... & 0 & 0\\
2j & 0 & 0 & ... & 0 & 0\\
0 & \small\mbox{$2j-1$}& 0 & ... & 0 & 0\\
... & ...& ... & ... & ... & ...\\
0 & 0 & 0 & ... & 0 & 0\\
0 & 0 & 0 & ... & 1 & 0
\end{array}
\right],\nonumber\\[3mm]
&& T^0 =\left[
\begin{array}{cccccc}
\small\mbox{$j$} & 0 & 0 & ... & 0 & 0\\
0 & \small\mbox{$j-1$} & 0 & ... & 0 & 0\\
0 & 0& \small\mbox{$j-2$} & ... & 0 & 0\\
... & ...& ... & ... & ... & ...\\
0 & 0 & 0 & ... & \small\mbox{$-j+1$} & 0\\
0 & 0 & 0 & ... & 0 &\small\mbox{$-j$}
\end{array}
\right].
\label{nine}
\end{eqnarray}
As was explained in Sect. 2, the algebraic sector of the periodic Lam\'e
system (\ref{lame})  consists of $2J+1$ eigenvalues, which are the
edges of the allowed bands or gaps. To relate the parameter $j$ in
(\ref{four}) to the QES parameter $J$ in (\ref{lame}) and (\ref{ham}) we
need to distinguish between $J$ even and $J$ odd.

\subsection{$J$ even}
\label{Jeven} 

For what follows it is convenient to introduce the operator
\beq
H  \equiv   -\frac{d^2}{d\phi^2} +4j(4j+1)\,\nu\,{\rm sn}^2(\phi|\nu )-  
2j(4j+1) \,.
\eeq
In this section we  consider four $sl(2)$ algebraizations
yielding the band boundaries for even values of $J$.
If  $J$ is {\em even}, then $J/4$ is semi-integer. As we
will see, in fact, in this case the 
full algebraic problem is split into four completely disconnected distinct
problems as follows:
\beq
2J+1 = \left\{2\left(\frac{J}{4}
\right) +1 \right\}  +  3\times\,  \left\{ 2\left(\frac{J-2}{4} 
\right) +1  \right\}  \,.  
\label{eight}
\eeq
We will consider four distinct algebraizations, one with $j= J/4$ and three with
$j= \frac{J}{4}-\frac 1 2$. The first two groups in Eq. (\ref{eight})
correspond to symmetric wave functions while the third and the fourth
to antisymmetric wave functions.

\subsubsection{ $j=J/4$, periodic wave function}
\label{4swf}

After the change of variables $\phi \to \eta (\phi )$,
the Hamiltonian on the left-hand side of (\ref{lame}) takes the form
\begin{eqnarray}
&&H= H \left( j\right) \equiv  -4\left[ (1-\nu )\eta -(1-2\nu
)\eta^2-\nu\eta^3
\right] d_\eta^{\,2} 
\nonumber\\
&&-2\left[(1-\nu ) -2(1-2\nu )\eta -3\nu\eta^2
\right] d_\eta +4j(4j+1)\,\nu\, (1-\eta) -2j(4j+1)\nonumber\\[4mm]
&&=-4\, \nu\,T^+T^0 - 4(1-2\nu ) T^+T^- - 4(1-\nu ) T^0T^-\nonumber\\[2mm]
&&-2\,\nu\, (6j+1)T^+ +4\, (2j+1)(1-2\nu ) T^0 -2\, (2j+1)(1-\nu )
T^-\nonumber\\[2mm]
&&+2j (1-2\nu )\,,
\label{twelve}
\end{eqnarray}
where
$j=\frac{J}{4}\,.$
The operator $H(j)$ defined here
is a basic element of the construction to be presented below.

It is clear that the Hamiltonian $H\equiv H ( j=\frac{J}{4} )$ is
purely algebraic; it acts on
$\Psi (\nu ) = P_{2j} (\nu )$, where $P_{2j}$ is a generic notation for a
polynomial of degree $2j$. There are
$2j+1$ eigenfunctions of the type $\Psi (\nu ) =
P_{2j} (\nu )$. These eigenfunctions,
$\Psi (\phi )= P_{2j} (\nu (\phi ))$, are periodic in $\phi$.
Moreover, $2j+1$ eigenvalues are most conveniently found
from the matrix realization (\ref{nine}), where $j$ is taken as $J/4$;
the dimension of the matrices $T^{\pm ,0}$ is $(2j+1)\times (2j+1)$. This
procedure gives the first contribution to the decomposition in
(\ref{eight}).

\subsubsection{$\tilde{j}=\frac{J-2}{4} $, periodic  wave function}
\label{perwafu}

It is known in the literature that the $sl(2)$ algebraization
of the problem (\ref{lame}) admits more than one solution for the
quasiphase $e^{-a}$. In Sect. \ref{4swf} the quasiphase was trivial, $e^{-a} = 1$.
Now we choose another solution, $e^{-a} = \sqrt{\eta (1-\eta )}$.
In other words,
\beq
\Psi (\eta ) = \sqrt{\eta (1-\eta )} \, \, \tilde\psi (\eta )\,.
\eeq
The quasigauge transformed Hamiltonian
\beq
H_{``G"}\equiv e^{ a} H e^{-a}
\label{thirteen}
\eeq
 acts on $\tilde\psi (\eta )$. Then
\begin{eqnarray}
H_{``G"} = H(\tilde j)-8\nu T^+ +4(2-3\nu)T^0 -4(1-\nu) T^-
+\left\{ 1+\nu+4\nu\, \tilde{j}\,\right\} \,,
\label{sixteen}
\end{eqnarray}
where
\beq
\tilde{j} =\frac{J}{4}-\frac{1}{2}
\eeq
and the generators $T^{\pm,0}$ on the right-hand side of Eq. (\ref{sixteen}),
including those in $H(\tilde j)$,
are in the representation $2\tilde{j} +1$ (i.e., the matrices (\ref{nine})
of dimension $(2\tilde{j} +1)\times (2\tilde{j} +1)$). 

We have to explain why the   wave function $\Psi (\phi )$  is periodic 
in this case. The situation is slightly more subtle than that in Sec.
\ref{4swf}.
The eigenfunctions have the form
\beq
\Psi = \sqrt{\eta   (1-\eta   )} \,P_{2\tilde j}(\eta )\,,
\eeq
where $\eta = \eta (\phi )$. Inside each  $\phi$ period the expression under the
square root touches zero twice; at these points care should be taken of the branches
of the square root. Needless to say that $\Psi (\phi )$ must be a smooth function of
$\phi$. Let us examine what happens, for instance, at $\phi$ near zero.
It this point $1-\eta = \phi^2$ and $\sqrt{ 1-\eta  }$ must be understood
as $\phi$; the square root is positive  at $\phi > 0$ and negative at
$\phi < 0$. This means that in the plane $\eta$ we pass from one branch of the
square root to another. This introduces a change of sign. Since this change happens
twice  inside the  $\phi$ period, the wave function $\Psi (\phi )$ we deal with is
periodic. If the change occurred once, the wave function would be
antiperiodic. This is case for two algebraizations considered in the next
subsection.

\subsubsection{$\tilde{j}=\frac{J-2}{4}$, antiperiodic wave functions}
\label{antiperwafu}

There are two more solutions for the quasiphase $e^{-a}$,
namely
\begin{eqnarray}
\Psi &=& \sqrt{(1-\eta)[(1-\nu)+\nu\,\eta ]}\,\,\tilde \psi
(\eta)\,,\nonumber\\[3mm]
 H_{``G"} &=&  H(\tilde j)-8\nu\,  T^+ +4\, (1-2\nu )\, T^0 -2\, (1-2\nu)\,
(1+2\tilde
j)\,,
\label{nineteen}
\end{eqnarray}
and
\begin{eqnarray}
\Psi &=& \sqrt{\eta\, [(1-\nu)+\nu\,\eta ]}\,\,\tilde \psi (\eta)\,,\nonumber\\[3mm]
H_{``G"} &=&  H(\tilde j)
-8\nu\,  T^+ +4\, (1-3\nu )\, T^0 -4\, (1-\nu )\,T^- 
\nonumber\\[3mm]
&-& 2\,  
(1+2\tilde j) +\nu\, (1+4\tilde j) \,.
\label{twenty}
\end{eqnarray}
In both cases the expression under the square root touches zero once
inside each  $\phi$ period. At the point where it occurs one passes from one
branch of the square root to another; correspondingly $\Psi (\phi )$ in the cases at
hand is antiperiodic. The generators $T^{\pm,0}$ in Eqs. (\ref{nineteen}),
(\ref{twenty}) are in the representation $2\tilde{j} +1$, while the eigenfunctions
$\tilde{\psi} (\eta )$ are polynomials of $\eta$ of degree
$2\tilde{j} $.  This concludes our consideration of the band boundaries for 
even values of $J$. 

\subsection{$J$ odd}
\label{Jodd} 

For what follows it is convenient to introduce the operator
\begin{eqnarray}
{\cal H} (j) 
&=&-4\, \nu\,T^+T^0 - 4(1-2\nu ) T^+T^- - 4(1-\nu ) T^0T^-
\nonumber\\[3mm]
&-&6\,\nu\, \left(2j+1\right) \, T^+ 
-4 (-1 - 2 j + 3 \nu + 4 j \nu)
\, T^0 -2\,   (1-\nu
)\, (1+2\,j)\,  T^-\nonumber\\[2mm] 
&+&  \nu -1 - 2 j  \,.
\label{twentytwo}
\end{eqnarray}

In this section we will consider four $sl(2)$ algebraizations
yielding the band edges for odd values of $J$.
If  $J$ is {\em odd}, then $(J-1)/4$ is semi-integer. As we
will see,  in this case the 
full algebraic problem is split in four  disconnected distinct
problems of the  following dimensions:
\beq
2J+1 = 3\times \left\{2\left(\frac{J-1}{4} 
\right) +1 \right\}  +   \left\{ 2\left(\frac{J-1}{4}-\frac{1}{2}
\right) +1  \right\}\, .  
\label{twentyp}
\eeq
 The first two groups in Eq. (\ref{twentyp})
correspond to symmetric wave functions while the third and the fourth
to antisymmetric ones.

\subsubsection{ $j=(J-1)/4$, periodic wave function} 
\label{pwfu}

The wave functions $\Psi (\phi)$ are periodic;
they have the structure
\beq
\Psi =\sqrt{[(1-\nu) +\nu\eta]}\,\,\tilde\psi (\eta )= \sqrt{[(1-\nu)
+\nu\eta]}\,\,P_{2j} (\eta)\,.
\eeq 
The quasigauge-transformed Hamiltonian acting on $\tilde\psi (\eta )$ 
has the form
\beq
H_{``G"} = {\cal H} (j) \,,
\label{twenty3}
\eeq
with the generators acting in the representation of dimension $2j+1$.

\subsubsection{ $\tilde{j}=(J-3)/4$, periodic wave function}
\label{pwfunction} 

There is another solution for the quasiphase leading to a periodic wave
function,
\begin{eqnarray}
\Psi &=&\sqrt{\eta\,(1-\eta )\, [(1-\nu) +\nu\eta]}\,\,\tilde\psi (\eta )
\nonumber\\[2mm]
&=&
\sqrt{\eta\,(1-\eta )\, [(1-\nu) +\nu\eta]}\,\, P_{2\tilde j} (\eta)\,,
\end{eqnarray}
where 
$ P_{2\tilde j} (\eta)$ is a polynomial of degree $2\tilde j$,
$$
\tilde j = j-\frac{1}{2} = \frac{J-3}{4}\,.
$$
In this case 
\begin{eqnarray}
H_{``G"} &=& {\cal H} (\tilde j) - 8\,\nu\, T^+  + 4\,(2-3\,\nu )\, T^0
-4\,(1-\nu )\, T^-\nonumber\\[2mm]
&+&   3 \nu + 4 \,\nu\, \tilde{j}-1
\,,
\label{twenty5}
\end{eqnarray}

\subsubsection{ $j=(J-1)/4$, antiperiodic wave functions}
\label{antipwf} 

The quasigauge factors emerging in two extra algebraizations leading to the
antiperiodic
$\Psi (\phi)$ are
$$
e^{-a} =\sqrt{\eta}\qquad\mbox{and}\qquad \sqrt{1-\eta}\,.
$$
This implies the following wave functions and quasigauge-transformed
Hamiltonians:
\begin{eqnarray}
\Psi &=& \sqrt{\eta} \,\,\tilde\psi (\eta )\,,
\nonumber\\[2mm]
H_{``G"} &=& {\cal H} ( j) +(1-\nu)\,(1+4\,j) + 4(1-\nu)\,T^0
- 4(1-\nu)\,T^-\,,
\label{twenty6}
\end{eqnarray}
and
\begin{eqnarray}
\Psi &=& \sqrt{1-\eta} \,\,\tilde\psi (\eta )
\nonumber\\[2mm]
H_{``G"} &=& {\cal H} ( j) +( 1+4\,j) + 4\,T^0\,,
\label{twenty7}
\end{eqnarray}
where $\tilde\psi (\eta )$ are polynomials of degree $2j$.

\section{Particular examples}
\label{PE}

It is instructive to consider a few simple examples.
This consideration will help us to establish a general pattern of the band edge
levels.

\subsection{$J =2$, five band-edge levels}
\label{J2}

The band (gap) edges in this case have been already presented in Eq.~(\ref{egs}).
According to the results of Sec. \ref{Jeven},
the QES sector consists of one doublet ($j=1/2$)  and three singlets
($\tilde{j}=0$). The doublet is obtained from Eq. (\ref{twelve}); the corresponding
energy eigenvalues are
$E_1$ and $E_5$.
Three singlets ($E_2,\,\,E_3$ and $E_4$) are obtained from Eqs.
(\ref{sixteen}), (\ref{nineteen}) and (\ref{twenty}).
 The doublet energies $E_1$ and $E_5$
are dual to each other; as far as the
singlets are concerned $E_4$ is dual to $E_2$
 ($E_4$ corresponds
to a periodic wave function while $E_2$ to an antiperiodic
wave function); and $E_3$ is self-dual, see Fig.
\ref{ppf3}.

\subsection{$J =3$, seven band-edge levels}
\label{J3}

\begin{figure}[htb]
\includegraphics[width=15cm]{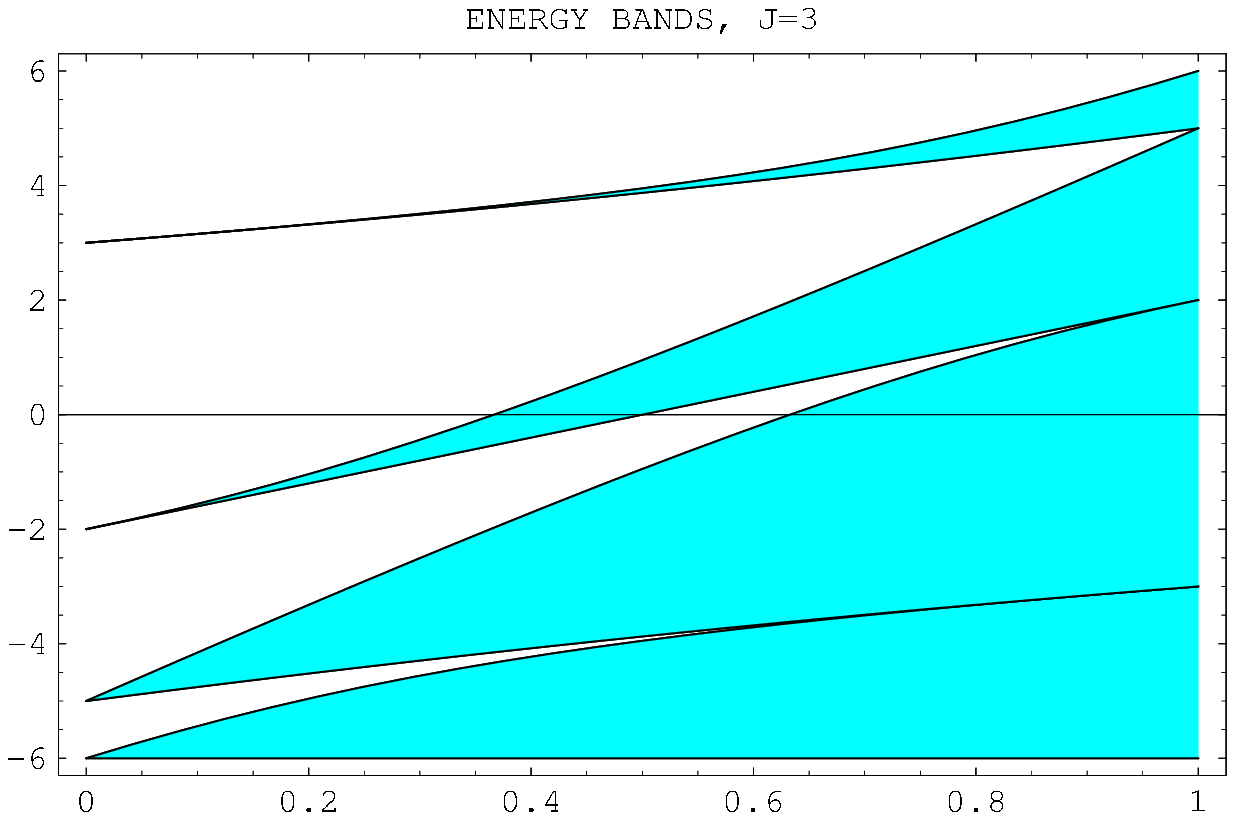}
\caption{
The energy bands for the Lam\'e system (\ref{lame}), as a
function of
$\nu$. The plot given is for
$J=3$. The first (from the bottom) and the third  allowed bands (unshaded) are
of the p-ap type, while the second and the fourth ones of the ap-p type, see text.}
\label{ppf4}
\end{figure}

According to the results of Sec.
\ref{Jodd}, 
the QES sector consists of three doublets ($j=1/2$) and
one singlet ($\tilde{j}=0$).
The doublet levels are obtained from Eqs. (\ref{twenty3}), (\ref{twenty6}),
and  (\ref{twenty7}),
\begin{eqnarray}
E_5&=&   -4  + 5\nu +2\sqrt{1 - \nu + 4\, \nu^2}\,, 
\nonumber
\\[2mm]
E_1 &=&   -4  + 5\nu -2\sqrt{1 - \nu + 4\, \nu^2}\,, 
\label{thirtytwop}
\end{eqnarray}
and
\begin{eqnarray}
E_6&=&   -1 + 2\nu + 2\sqrt{4 - \nu + \nu^2}\,,
\nonumber\\[2mm]
E_2 &=&   -1 + 2\nu - 2\sqrt{4 - \nu + \nu^2}\,,
\label{thirtythree}
\end{eqnarray}
 and
\begin{eqnarray}
E_7&=&  -1 + 5\nu + 2\sqrt{4 - 7\nu + 4\, \nu^2}\,,
\nonumber
\\[2mm]
E_3 &=&   -1 + 5\nu - 2\sqrt{4 - 7\nu + 4\, \nu^2}\,.
\label{thirtyfourp}
\end{eqnarray}
The singlet is   obtained from Eq. (\ref{twenty5}),
\beq
E_4 =  - 2(1-2\nu)\,.
\label{thirty7}
\eeq
The doublet energy eigenvalues 
$E_2$ and $E_6$ are dual to each other,
$E_1$ and $E_7$ are dual, and so are $E_5$ and $E_3$.
The singlet energy eigenvalue $E_4$ is self-dual, see Fig.~\ref{ppf4}.

\subsection{$J =4$, nine band-edge levels}
\label{nineband}

According to the results of Sec. \ref{Jeven},
the QES sector consists of one triplet ($j=1 $)  and three doublets
($\tilde{j}=1/2$). The triplet is obtained from Eq. (\ref{twelve}),
\begin{eqnarray}
E_{1,5,9}&=&  -\frac{10}{3}\, (1-2\nu )
-\frac{8\sqrt{13}}{3}\,\sqrt{1-\nu+\nu^2}\, \cos\left[\delta_{1,5,9} \right.
\nonumber\\[3mm]
&-&\left.
\!\! \!\! \frac{1}{3}\,{\rm arcsin}
\frac{3\sqrt{3}\,\sqrt{144-432\nu
+2089\nu^2-3458\nu^3+2089\nu^4-432\nu^5+144\nu^6
}}{26\sqrt{13}\, (1-\nu+\nu^2)^{3/2}}\,\,
\right],
\nonumber\\[5mm]
\delta_1&=&\frac{\pi}{3}\,,\qquad\delta_5=-\frac{\pi}{3}\,,\qquad
\delta_9=\pi \,.
\label{thirtyeight}
\end{eqnarray}
At $\nu < 1/2$, the arcsine on the right-hand side is in the first quadrant,
and is defined in a standard way. At  $\nu > 1/2$ it must be defined as a smooth
analytic continuation.

\begin{figure}[htb]
\includegraphics[width=15cm]{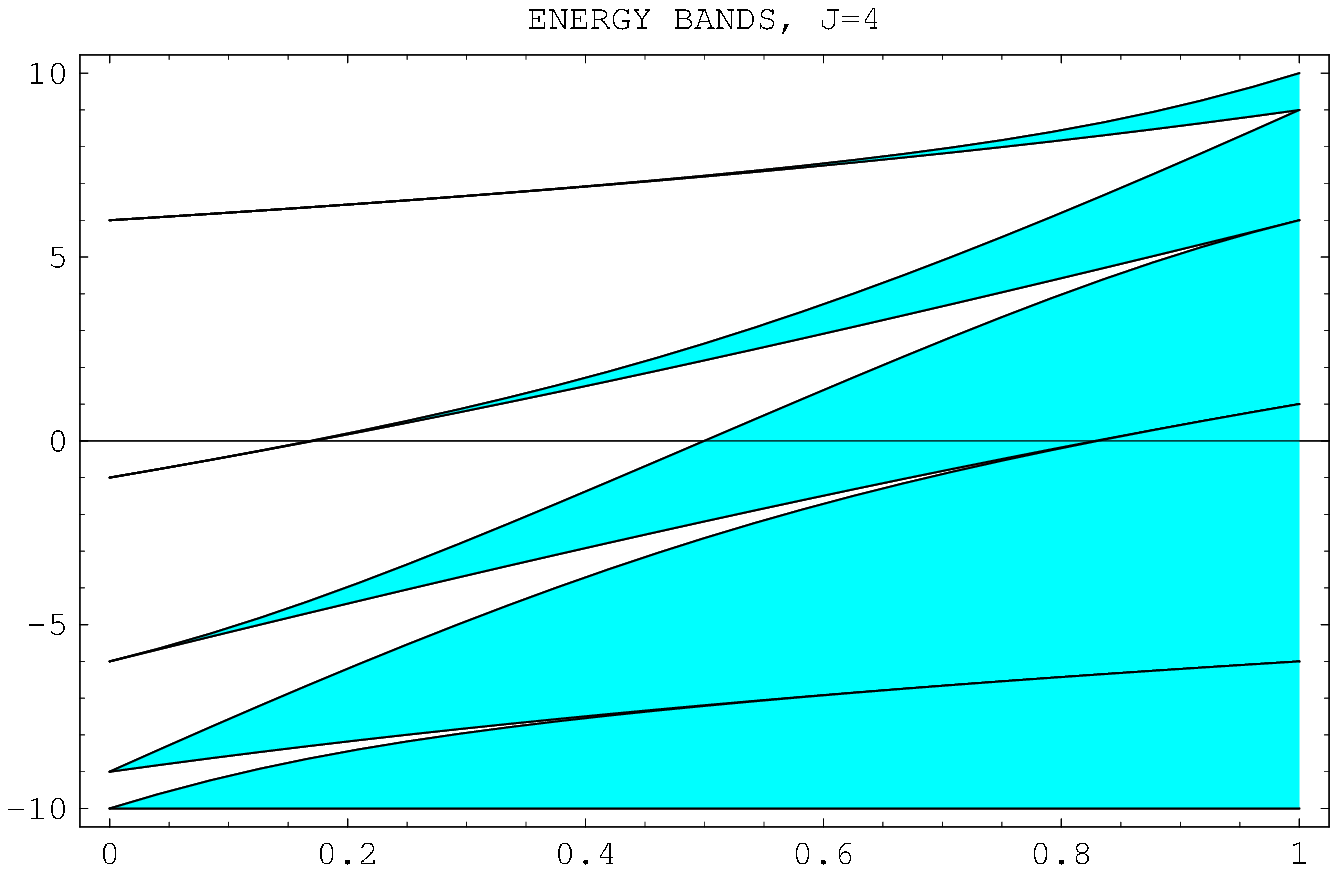}
\caption{The energy bands for the Lam\'e system (\ref{lame}), as a
function of $\nu$. This is the plot for
$J=4$. The first (the lowest), the third and the fifth allowed bands (unshaded) are
of the p-ap type, while the second and the fourth of the ap-p type, see text.}
\label{ppf1}
\end{figure}

The doublet levels are obtained from Eqs. (\ref{sixteen}), (\ref{nineteen})
and (\ref{twenty}),
\begin{eqnarray}
E_4&=& 5\nu - 2\sqrt{9 - 9\nu + 4\nu^2}\,, 
\nonumber
\\[2mm]
E_8 &=&    5\nu + 2\sqrt{9 - 9\nu + 4\nu^2}\,, 
\label{fourty}
\end{eqnarray}
and
\begin{eqnarray}
E_3&=&    -5+10\nu - 2\sqrt{4 - 9\nu + 9\nu^2}\,,
\nonumber\\[2mm]
E_7 &=&   -5+10\nu + 2\sqrt{4 - 9\nu + 9\nu^2}\,,
\label{fourtyone}
\end{eqnarray}
 and
\begin{eqnarray}
E_2&=&   -5+5\nu - 2\sqrt{4 +\nu + 4\nu^2}\,,
\nonumber
\\[2mm]
E_6 &=&  -5+5\nu + 2\sqrt{4 +\nu + 4\nu^2}\,.
\label{fourty3}
\end{eqnarray}
The triplet levels (\ref{thirtyeight}) are  dual, and so are
the doublet levels $E_3$ and $E_7$. Moreover, the
two remaining doublets are dual to each other, namely
$E_4$ is dual to
$E_6$, while $E_8$ to $E_2$, see Fig. \ref{ppf1}.

\subsection{$J =5$, eleven band-edge levels}
\label{elevenband}

According to the results of Sec.
\ref{Jodd} 
the QES sector consists three triplets ($j=1$) and
one doublet ($\tilde{j}=1/2$).

The triplet levels are obtained from Eqs. (\ref{twenty3}),  (\ref{twenty6})
and  (\ref{twenty7}). It is not difficult  to obtain that
\begin{eqnarray}
E_{1,5,9} &=&  \frac{5}{3}\, (-5 + 7 \nu) -\frac{8}{3}\,\sqrt{13 -
13\nu + 28\nu^2}\,
\cos \Big[ \delta_{1,5,9}  
\nonumber\\[3mm]
&-&\left.
\!\! \!\! \frac{1}{3}\,{\rm arcsin} \frac{9\sqrt{3}\,
    \sqrt{16 - 48\nu + 461\nu^2 - 842\nu^3 + 541\nu^4 - 128\nu^5 + 
    256\nu^6}}{2 (13 - 13\nu + 28\nu^2)^{3/2}} \,\,\right],
\nonumber\\[5mm]
\delta_1&=&\frac{\pi}{3}\,,\qquad\delta_5=-\frac{\pi}{3}\,,\qquad
\delta_9=\pi \,.
\label{fourty5}
\end{eqnarray}
This triplet is dual to that of Eq. (\ref{fourty7}). As in Eq. (\ref{thirtyeight}), the
arcsine on the  right-hand side  is defined in the standard way when it is in the
first quadrant, (i.e. at $\nu <\nu_*= 0.371...$) while at  larger $\nu$
the
arcsine is understood as a smooth analytic continuation \footnote{In Eqs.
(\ref{fourty6}) and (\ref{fourty7}) $\nu_*=1/2$ and $0.629..$, respectively.}.

Furthermore,
\begin{eqnarray}
E_{2,6,10} &=&  -\frac{10}{3}\, (1-2\nu )
-\frac{8}{3}\,\sqrt{28 - 13\nu + 13\nu^2}\, \cos\Big[\delta_{2,6,10}  
\nonumber\\[3mm]
&-&\left.
\!\! \!\! \frac{1}{3}\,{\rm arcsin}
\frac{9\sqrt{3}\,\sqrt{256 - 128\nu + 541\nu^2 - 842\nu^3 + 461\nu^4 - 48\nu^5 + 
    16\nu^6
}}{2 \, (28 - 13\nu + 13\nu^2)^{3/2}}\,\,
\right],
\nonumber\\[5mm]
\delta_2 &=&\frac{\pi}{3}\,,\qquad\delta_6 = -\frac{\pi}{3}\,,\qquad
\delta_{10}=\pi \,.
\label{fourty6}
\end{eqnarray}
This triplet of levels is dual to itself.

Finally, the third triplet is
\begin{eqnarray}
&&E_{3,7,11}=  \frac{5}{3}\, (-2 + 7 \nu) -\frac{8}{3}\,\sqrt{28 -
43\nu + 28\nu^2}\,
\cos \Big[ \delta_{3,7,11}  
\nonumber\\[3mm]
&&-\,\, \left.
\!\! \!\! \frac{1}{3}\,{\rm arcsin} \frac{9\sqrt{3}\,
    \sqrt{ 256 - 1408\nu + 3741\nu^2 
-5162\nu^3 + 3741\nu^4 - 1408\nu^5 
 +256 \nu^6}}{2 (28 - 43\nu +
28\nu^2)^{3/2}} \,\right],
\nonumber\\[5mm]
&&\delta_3  = \frac{\pi}{3}\,,\qquad\delta_7=-\frac{\pi}{3}\,,\qquad
\delta_{11}=\pi \,.
\label{fourty7}
\end{eqnarray}
It is dual to the triplet (\ref{fourty5}).

\begin{figure}[htb]
\includegraphics[width=15cm]{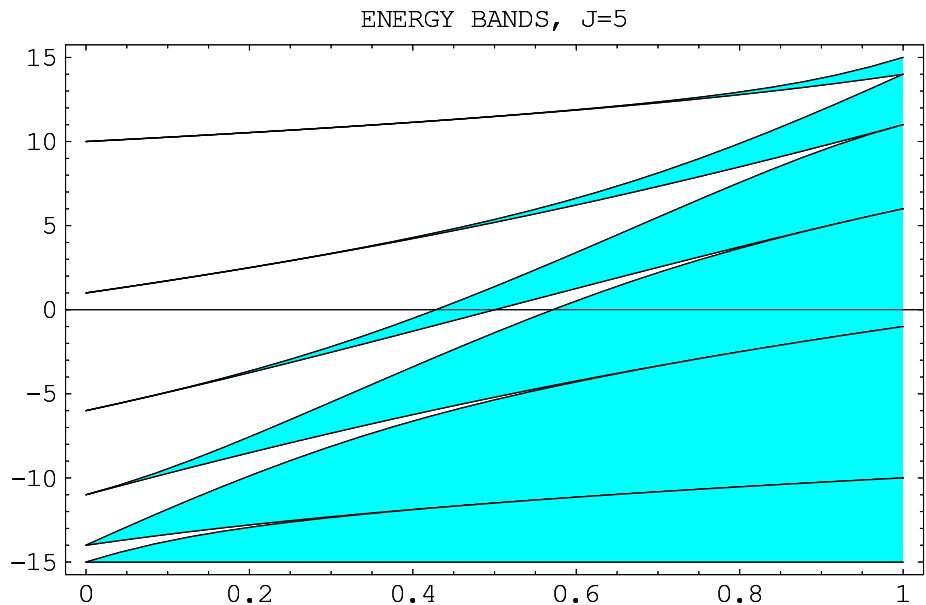}
\caption{The energy bands for the Lam\'e system (\ref{lame}), as a
function of $\nu$. This is the plot for
$J=5$. The first (the lowest), the third and the fifth allowed bands (unshaded) are
of the p-ap type, while the second, the fourth and the sixth of the ap-p type, see
text.}
\label{ppf2}
\end{figure}

The doublet levels are obtained from Eq. (\ref{twenty5}), 
\begin{eqnarray}
E_4&=&  -5  + 10\nu - 6\sqrt{1 - \nu + \nu^2}\,, 
\nonumber
\\[2mm]
E_8&=&    -5  + 10\nu + 6\sqrt{1 - \nu + \nu^2}\,.
\label{fourty9}
\end{eqnarray}
These two eigenvalues are dual to each other. The overall band structure
for $J=5$  is presented in Fig. \ref{ppf2}.

\subsection{The general pattern}
\label{pattern}

\begin{figure}[htb]
\includegraphics[width=15cm]{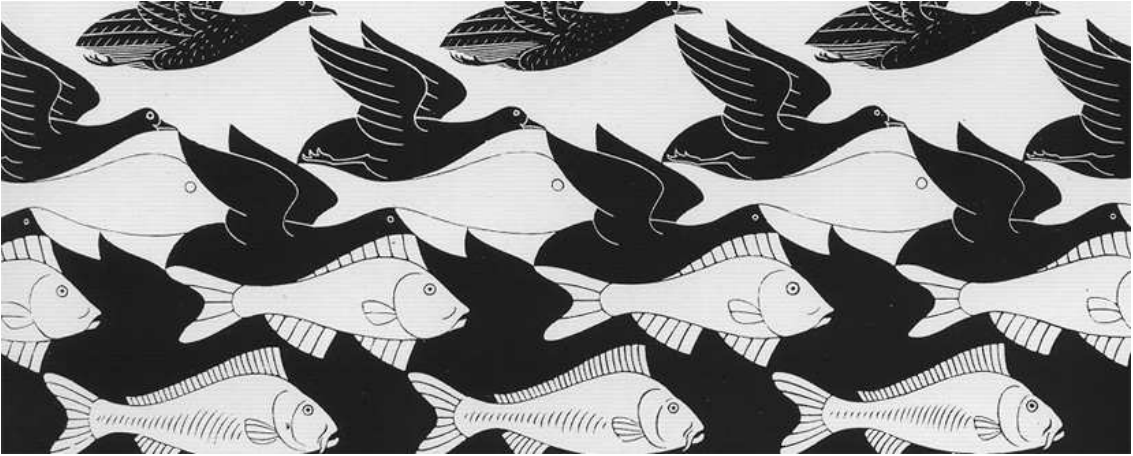}
\caption{An illustration of duality: a fragment of Escher's ``Sky and
Water."}
\label{ppf5}
\end{figure}

The remarkable features of duality and energy-reflection symmetry of the spectral
problem (\ref{lame}) reveal  themselves  in a transparent manner  in
Figs.~\ref{ppf3},
\ref{ppf4}, \ref{ppf1} and \ref{ppf2}.
The allowed bands of the elliptic potential $\nu\, {\rm sn}^2(\phi|\nu )$
present (up to a sign) the forbidden bands (gaps) of the potential
$(1-\nu)\, {\rm sn}^2(\phi|1- \nu )$, and {\em vice versa}. A nice illustration for   duality
of this type was given long ago by M. Escher.
Figure \ref{ppf5} shows a fragment of Escher's ``Sky and Water." Here
the gaps between the birds are fishes, while the gaps between the fishes 
are birds.

The last allowed band extends up to $E=\infty$.
The lower boundary of this band corresponds to periodic wave functions for
even values of $J$, and to anti-periodic ones for odd values of $J$.
The p-ap and ap-p allowed bands alternate, and so do the gaps. At $\nu=1/2$,
the energy bands are self-dual. This means that for each band (gap) edge
with the energy $E$ there is one with the energy $-E$.  

As $\nu\to 1$, the
widths of the allowed bands  (except the last one) tend to
zero. In fact, these bands shrink to the discrete bound states of the P\"oschl-Teller
system. The lowest band is at $E=-J(J-1)/2$, the last band starts at $E=J(J+1)/2$.
On the contrary, the gap widths decrease. The width of the lowest gap is
$\Delta E = 2J-1$, and then it decreases: $\Delta E = 2J-3,\,\,\, 2J-5,...., 1$. 

As $\nu\to 0$, the gap widths tend to zero, while the allowed band widths
grow. The lowest allowed band is at
$-J(J+1)/2\leq E\leq -J(J+1)/2 -1$, its width is $\Delta E =  1$, the next band is at
$-J(J+1)/2 -1\leq E\leq -J(J+1)/2 -4$, its width is $\Delta E =  3$, and so on.
The last (infinite-width) allowed band starts at $E=J(J-1)/2$.
The bottom edge of the last (infinite-width) allowed band is of the p type
for  even $J$ and ap type for
odd $J$. 

In each $(2j+1)$-plet ($j>0$) of the $sl(2)$ representation (Sect. \ref{QES})
the energy level pattern is as follows:
$$
E_{k_0}\,,\,\,\, E_{k_0 +4}\,,\,\,\, E_{k_0 +8}\,, \dots\,.
$$
The multiplet which includes $E_{J+1}$ is special:
it is always self-dual. The $(J+1)$-th energy eigenvalue
$E_{J+1}$ is dual to itself and passes through zero at $\nu=1/2$. 
The dimension of this multiplet is $2\, [J/4] + 1$, where
[...] denotes the entire part.

The pattern described above is readily understood
from the representation (\ref{ham}) and/or the
$sl(2)$-based representations considered in Sect.~\ref{QES}.

\section{Energy reflection symmetry at the self-dual \\
point $\nu =1/2$.}
\label{nu=1/2}

When $\nu=\frac{1}{2}$, the Lam\'e system is self-dual and so has an
energy-reflection (ER) symmetry. This symmetry implies that each
eigenlevel of energy $E$ is accompanied by a level of energy
$-E$, the corresponding wave functions being related in a well-defined manner.
A class of ER symmetric  QES problems 
was constructed in \cite{st2}. 
Although the construction of Ref. \cite{st2} guarantees ER-symmetric
spectra, by no means does it present a sufficient condition for the ER
symmetry. The spectral problem (\ref{lame}), with $\nu =1/2$,
is in fact an  expansion of the class of the ER-symmetric  problems  (the
mechanism leading to this  expansion, the possibility of different
algebraizations of one and the same spectral problem
due to the existence of distinct solutions for the quasi-gauge, was overlooked in
\cite{st2}). The focus of this section  is
the ER symmetry properties of periodic elliptic  potentials from the
standpoint of QES, and their connection with the Lam\'e system
(\ref{lame}).

To begin, we observe that at $\nu =1/2$ the function sn$^2 (\phi
|\nu )$ is related to the Weierstrass function ${\cal P}$ with the
invariants \cite{abramowitz}
\beq
g_2 =4\,,\qquad g_3 =0\,;
\label{fifty}
\eeq
namely, 
\beq
\frac{J(J+1)}{2}\, \left\{ 
{\rm sn}^2(\phi|\nu )- 1
\right\} \equiv - \, \frac{J(J+1)}{2}\,\, \frac{1}{{\cal P}\left(\frac{\phi}{\sqrt 2}-
\frac{\tau}{ 2}
\right)},
\eeq
where
\beq
\tau = \frac{2\sqrt{\pi}\Gamma (5/4)}{\Gamma (3/4)}
\label{tauperiod}
\eeq
is the period of the Weierstrass function ${\cal P} (x;\,  g_2 =4, g_3 =0)$. The invariants
$g_{2,3}$ will be suppressed below. This relation implies, in turn, that at $\nu =1/2$
the spectral problem (\ref{lame}) can be written as
\beq
\left\{ -\frac{1}{2}\, \frac{d^2}{dx^2} - \frac{J(J+1)}{2}\,\,
\frac{1}{{\cal P}\left(x -
\frac{\tau}{ 2}
\right)}
\right\}
\Psi (x) = E\, \Psi (x)\,,
\label{1/2}
\eeq
where the following change of variables is made:
\beq
\frac{\phi}{\sqrt 2}\to x\,.
\eeq
Note the occurrence of the factor $1/2$ in the kinetic term. Weierstrass 
function related  potentials  
first surfaced in the context of the Lie-algebraic construction in 
\cite{turbiner,turbiner2}, and are well known in the general study of
finite-gap potentials \cite{belokolos}.

\begin{figure}[htb]
\includegraphics[width=15cm]{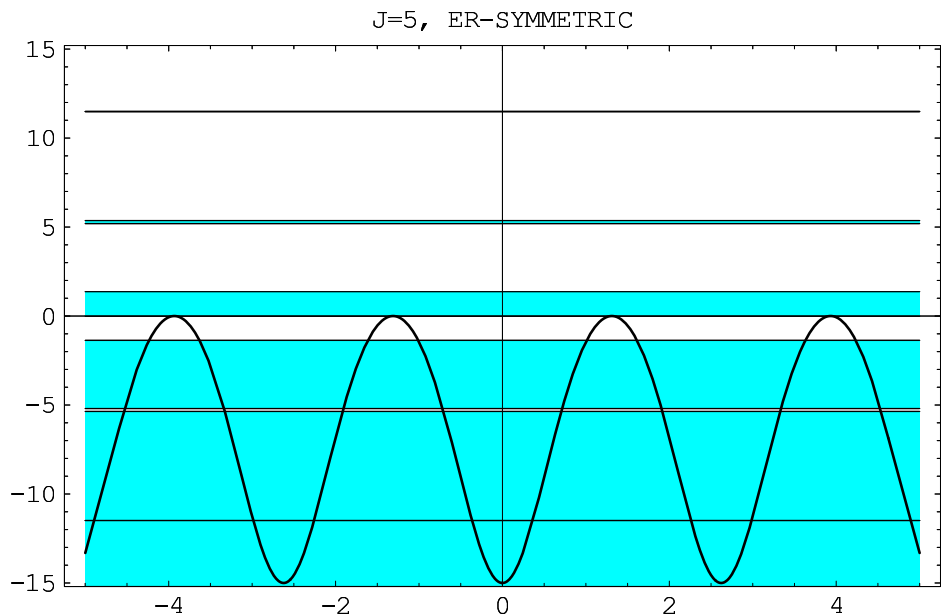}
\caption{The ER-symmetric potential (\ref{1/2}) versus $x$
for $J=5$. The gaps are shaded while the allowed bands are unshaded.
Notice the ER symmetry which interchanges bands and gaps under $E\to -E$.}
\label{ppf7}
\end{figure}

In the general case of sn$ (\phi |\nu )$ the parallelogram of periods is a rectangle; the
$\nu=1/2$ case is special: the parallelogram of periods is a square. This
is called the "lemniscate" case \cite{abramowitz}. The auxiliary variable
$\eta (\phi) = 1 - {\rm sn}^2(\phi|\nu )$ (see Eq. (\ref{threes})) becomes
\beq
\eta (x) \equiv \frac{1}{{\cal P}(x)}\,.
\eeq
The analogue of Eq. (\ref{twos}) becomes
\beq
\left(\frac{d\,\eta}{d\,x}
\right)^2 =  4(\eta - \eta^3)\,.
\label{fifty6}
\eeq
A crucial relation
\beq
\eta (ix) = -\eta (x)
\eeq
follows from the properties of the Weierstrass function
with the invariants (\ref{fifty}).
For real $x$ the function $\eta (x)$ varies in the interval
$[0,1]$; it is doubly periodic in the complex plane
(along the real and imaginary axes), with periods
$\tau$ and $i\tau$. An example of the ER-symmetric potential (\ref{1/2})
($J=5$), with the corresponding band structure, is
shown in Fig.
\ref{ppf7}.

\subsection{J even, self-dual level multiplets}
\label{Jevenself}

 Two solutions of this type for the band edges
(i.e. individually ER-symmetric)
were found some time ago in Ref. \cite{st2} (see Eq. (17)
with $\nu = 0$ and $1/2$). The first one (periodic eigenfunctions, cf. Eq. (\ref{twelve})) is
\begin{eqnarray}
j&=&\frac{J}{4}\,,\qquad \Psi (x) = P_{2j} (\eta (x ))\,,\\[0.2cm]
H &=& - 2 (\eta -\eta^3) d^2_\eta +(3\eta^2-1)d_\eta -2j (4j+1)\eta
\nonumber\\[0.2cm]
&=&-2T^+T^0 - 2 T^0T^- -(6j+1)T^+ -(2j+1)T^-\,.
\label{thur1}
\end{eqnarray}
The second solution (anti-periodic eigenfunctions, cf. Eq.
(\ref{nineteen})) is
\begin{eqnarray}
\tilde j&=&\frac{J-2}{4}\,,\qquad \Psi (x ) =\sqrt{1-[\eta (x)]^2} \, \,P_{2\tilde j} \, (\eta
(x ))\,,
\nonumber\\[0.2cm]
 H_{``G"} &=& - 2 (\eta -\eta^3) d^2_\eta +(7\eta^2-1)d_\eta -2\tilde j (4\tilde j+5)\eta
\nonumber\\[0.2cm]
&=&-2T^+T^0 - 2 T^0T^- -(6\tilde j+5)T^+ -(2\tilde j+1)T^-\,.
\label{thur2}
\end{eqnarray}
Note that the  ER symmetry of each of the two  level multiplets above
follows from the fact that 
$$ H\to - H$$ under the replacement $x\to ix,\,\,\,\eta \to - \eta$.
The corresponding wave functions are obtained from one another by
the replacement $\eta  \to - \eta $. This is the mechanism discussed in
\cite{st2}. In the example of Sec. \ref{nineband}, $J=4$,
the band edges under discussion are
$$
\{2\sqrt{13}\,,\,\,\, 0\,,\,\,\,-2\sqrt{13} \}\,,\qquad 
\{\sqrt{7},\,\,\, -\sqrt{7}\}\,.
$$

\subsection{J even, a level multiplet of the p type dual to that of 
the ap type}
\label{Jevendual}

Our assertion is: there exists another mode in which the ER symmetry 
can be realized in the problem (\ref{lame}). It does not fall into the
category specified by Eqs. (6) and (8) in Ref.
\cite{st2}. We consider two {\em distinct} quasigauge transformations,
both leading to Lie-algebraic $H_{``G"}$'s, such that the quasiphase factors
$e^{-a}$ are not invariant under $\eta\to-\eta$, but, rather, are transformed one into
another.  This will lead to a wider class of ER-symmetric problems than that
considered in Ref.
\cite{st2}. 
The  pair of  conjugate Hamiltonians is
\begin{eqnarray}
\tilde j&=&\frac{J-2}{4}\,,\qquad \Psi (x ) =\sqrt{\eta (x)\left\{
1-  \eta (x) \right\}} \,
\,
P_{2\tilde j} \, (\eta
(x ))\,,
\nonumber\\[0.2cm]
H_{``G"} &=& - 2T^+T^0 - 2 T^0T^- - 2\left( 3\tilde j+\frac{5}{2} \right)T^+
-\left( 2\tilde j+3\right)T^- 
\nonumber\\[0.2cm]
&+& 2T^0 + \left( 2\tilde j+\frac{3}{2} \right)\,,
\label{c3b}
\end{eqnarray}
(periodic eigenfunctions, cf. Eq. (\ref{sixteen})),
and
\begin{eqnarray}
\tilde j&=&\frac{J-2}{4}\,,\qquad \Psi (x ) =\sqrt{\eta (x)\left\{
1+  \eta (x) \right\}} \,
\,P_{2\tilde j} \, (\eta (x ))\,,
\nonumber\\[0.2cm]
H_{``G"} &=& - 2T^+T^0 - 2 T^0T^- - 2\left( 3\tilde j+\frac{5}{2} \right)T^+
-\left( 2\tilde j+3\right)T^-
\nonumber\\[0.2cm]
 &-& 2T^0 - \left( 2\tilde j+\frac{3}{2} \right)\,,
\label{c3a}
\end{eqnarray}
(anti-periodic eigenfunctions, cf. Eq. (\ref{twenty})).

The terms containing  $T^0$ would be forbidden by the ansatz considered in
Ref. \cite{st2}. However, it is clear that the Hamiltonians
(\ref{c3b}) and  (\ref{c3a}) swap under the substitution $\eta\to -\eta$. 
In the example of Sec. \ref{nineband}, $J=4$,
the band edges under discussion are
$$
\left\{ 
 \frac{5}{2}+\sqrt{22}\,,
\,\,\, 
 \frac{5}{2}-\sqrt{22}\right\},
\,\,\,
\mbox{periodic}
\,;
\qquad 
\left\{
-\frac{5}{2} -\sqrt{22}\,,
\,\,\,
-\frac{5}{2} +\sqrt{22} 
\right\},\,\,\,\mbox{anti-periodic}\,.
$$

\subsection{J odd, self-dual level multiplets} 
\label{Joddself}

 Two solutions of the {\em self}-conjugate type for the band edges  
are as follows:
\begin{eqnarray}
j&=&\frac{J-1}{4}\,,\qquad \Psi (x) = \sqrt{\eta (x) } \,\, P_{2j} \,
(\eta (x))\,,
\nonumber\\[2mm]
 H_{``G"} &=& - 2T^+T^0 - 2 T^0T^- - 2\left( 3j+\frac{3}{2}
\right)T^+ -\left( 2j+3\right)T^-   \,,
\end{eqnarray}
(anti-periodic eigenfunctions, cf. Eq. (\ref{twenty6})), 
and
\begin{eqnarray}
\tilde j &=& \frac{J-3}{4}\,,\qquad \Psi (x) = \sqrt{\eta (x) \left[1- (\eta (x))^2\right]} \,\,
P_{2\tilde j}
\, (\eta (x))\,,
\nonumber\\[2mm]
H_{``G"} &=& - 2T^+T^0 - 2 T^0T^- - 2\left( 3\tilde j+\frac{7}{2} \right)T^+
-\left( 2\tilde j+3\right)T^-   \,.
\end{eqnarray}
(periodic eigenfunctions, cf. Eq. (\ref{twenty5})).
These two solutions  fall into the category treated in  Ref. \cite{st2}. 
In the example of Sec. \ref{elevenband}, $J=5$,
the band edges under discussion are
\begin{eqnarray}
&&\{2\sqrt{33}, \,\,\, 0, \,\,\, -2\sqrt{33}\}, \,\,\, 
\mbox{anti-periodic};
\nonumber\\[2mm]
&&\{3\sqrt{3}, \,\,\, -3\sqrt{3}\}, \,\,\, \mbox{periodic}\,.
\nonumber
\end{eqnarray}

\subsection{J odd, a level multiplet of the p type dual to that of the ap
type}
\label{Jodddual}

The  pair of conjugate Hamiltonians is
\begin{eqnarray}
j &=& \frac{J-1}{4}\,,\qquad \Psi (x) = \sqrt{1+\eta (x) } \,\, P_{2j} \,
(\eta (x))\,,
\nonumber\\[2mm]
H_{``G"} &=& - 2T^+T^0 - 2 T^0T^- - 2\left( 3j+\frac{3}{2} \right)T^+
-\left( 2j+1\right)T^- 
\nonumber\\[2mm]
 &-& 2T^0  - \left( 2j+\frac{1}{2} \right)\,,
\label{c2a}
\end{eqnarray}
(periodic eigenfunctions, cf. Eq. (\ref{twenty3})),
and
\begin{eqnarray}
j &=& \frac{J-1}{4}\,,\qquad \Psi (x) = \sqrt{1- \eta (x) } \,\, P_{2j} \,
(\eta (x))\,,
\nonumber\\[2mm]
H_{``G"} &=& - 2T^+T^0 - 2 T^0T^- - 2\left( 3j+\frac{3}{2} \right)T^+
-\left( 2j+1\right)T^- 
\nonumber\\[2mm]
&+& 2T^0 + \left( 2j+\frac{1}{2} \right)\,,
\label{c2b}
\end{eqnarray}
(anti-periodic eigenfunctions, cf. Eq. (\ref{twenty7})). In the example of
Sec. \ref{elevenband}, $J=5$, the band edges under discussion are
\begin{eqnarray}
&&\left\{
-\frac{5}{2}-4\sqrt{6}\cos \left(\delta_{1,5,9}-\frac{\pi}{3}+
\frac{1}{3}
{\rm arcsin} \sqrt{ \frac{191}{216}}\right)
\approx  -11.49 ,\, -1.37,\, 5.36
\right\}, \,\,\, 
\mbox{p};
\nonumber\\[2mm]
&&\left\{
 \frac{5}{2}+ 4\sqrt{6}\cos \left(
\delta_{3,7,11}-\frac{\pi}{3}+
\frac{1}{3}
{\rm arcsin} \sqrt{ \frac{191}{216}}\right)
\approx -5.36,\, 1.37,\, 11.49
\right\},\,\,\, 
 \mbox{a-p} .
\nonumber
\end{eqnarray}

\section{Weak coupling versus quasiclassical expansion}
\label{nonpert}

In this section we describe how the duality transformation (\ref{duality})
connects various perturbative and nonperturbative techniques, by relating
information about states high up in the spectrum to states low down in the
spectrum. We first consider approximate techniques for the {\it locations}
of bands and gaps. Next we consider techniques for evaluating the {\it
widths} of bands and gaps. The calculations of widths are sensitive to
exponentially small contributions which are neglected in the calculations of
the locations.

\subsection{Estimates of locations of bands and gaps}
\label{locations}

Because of the duality transformation (\ref{duality}), the location of a
low-lying band in the spectrum is related to the location of a high-lying
gap in the dual spectrum (i.e., the spectrum for the potential obtained by
making the duality replacement $\nu\to 1-\nu$).  
The location of the low-lying bands can be obtained by applying the weak coupling
expansion. The location of a high-lying
gaps  can be obtained by applying the quasiclassical expansion.
In both cases the quadratic Casimir
$J(J+1)$ determines  the expansion parameter.
It is convenient to introduce a parameter $\kappa$,
\beq
\kappa =\sqrt{J(J+1)}\,.
\label{kappa}
\eeq
Then $1/\kappa$ is the weak coupling constant of the perturbative expansion.
Simultaneously, $1/\kappa$ plays the role of $\hbar$
in the quasiclassical expansion.

\subsubsection{Perturbation theory for location of lowest band}
\label{ptbandlocation}

In the limit $J\to\infty$, the width of the lowest band becomes very
narrow, so it makes sense to estimate the ``location" of the band. In fact, as
we will see in the next sections, the width  shrinks exponentially fast, so
we can estimate the location of the band to within exponential accuracy
using elementary perturbation theory. That is, for large $J$, we can consider
a single isolated well of the periodic Lam\'e potential, and expand \cite{abramowitz}
near $\phi =0$, keeping quartic, sextic and higher order anharmonic terms, 
\begin{eqnarray} 
{\rm sn}^2(\phi|\nu)=\phi^2-\frac{\nu+1}{3}\phi^4+\frac{2+13
\nu+2\nu^2}{45}\phi^6+\dots\,.
\label{sn2}
\end{eqnarray}
Rescaling the coordinate, 
$$\phi=     \frac{x}{(\nu J(J+1))^{1/4}}\equiv \nu^{-1/4}\,\kappa^{-1/2}\, x\,,$$ 
the Lam\'e equation
(\ref{lame}) becomes
\beq
 \left[-\frac{d^2 }{d x^2} +x^2
-\frac{\nu+1}{3\kappa \,\sqrt{\nu}} \,    x^4 + 
\frac{2+13\nu+2\nu^2}{45 \kappa^2 \,\nu }\, x^6
+\dots \right]\, \Psi  =
\frac{E+\frac{1}{2}\kappa^2}{\kappa\, \sqrt{\nu }}\,
\, \Psi \,.
\label{ptlame}
\eeq
Thus, the lowest energy level can be evaluated as a simple series in powers
of $1/\kappa$,
\beq
E_0 = -\frac{1}{2}\, \kappa^2 
\left[1-\frac{2\sqrt{\nu}}{\kappa}
+\frac{\nu+1}{2\, \kappa^2}+
\frac{(1-4
\nu+\nu^2)}{8\,\sqrt{\nu}\,\kappa^3} +O\left(\frac{1}{\kappa^4}\right)\right]\,.
\label{e0pt}
\eeq

\vspace{1mm}

\subsubsection{Semiclassical method for location of highest gap}
\label{wkbgaplocation}

As $J\to\infty$, the location of the highest gap can be found by semiclassical techniques. 
First, note that for a given $\nu$, as $J\to\infty$ the highest gap lies above the top of 
the potential. Thus, the turning points lie off the real $\phi$ axis. For a periodic
potential the gap edges occur when the discriminant \cite{magnus,peierls} 
takes values $\pm 1$.
By WKB, the discriminant is
\begin{eqnarray}
\Delta(E)=\cos \left( \frac{1}{\hbar}\,\,  \sum_{n=0}^\infty \, \hbar^n\,
S_n\, (P)\right)\,,
\label{disc}
\end{eqnarray}
where $P$ is the period, and the $S_n(x)$ are the standard 
WKB functions (see e.g. \cite{carlbook}),
\begin{eqnarray}
S_0(x) &=& \int^x_0 \sqrt{Q(t)}\, dt\,,\nonumber\\[2mm]
S_1(x) &=& -\frac{1}{4}\left[\log Q(x)\right]_0^x\,, \nonumber\\[2mm]
S_2(x) &=& 
\int^x_0 \left[\frac{Q^{\prime\prime}}{8 Q^{3/2}}-\frac{5(Q^\prime)^2}{32\,
Q^{5/2}}\right]\, dt \,,
\nonumber\\[2mm] 
S_3(x)&=& \left[-\frac{Q^{\prime\prime}}{16
Q^2}+\frac{5(Q^\prime)^2}{64\,  Q^3} \right]_0^x\,, 
\label{wkbs}
\end{eqnarray}
and so on. Here 
$$
Q(x)\equiv E-V(x)\,.
$$ 
These functions $S_n(x)$ can be generated to any order by a  simple
recursion formula \cite{carlbook}. The $J$-th gap occurs when the
argument of the cosine in the discriminant (\ref{disc}) is $J\pi$. This
gives the location of the $J$-th gap, neglecting exponential
corrections which correspond to the exponentially narrow width of the gap
in the semiclassical $J\to\infty$ limit.

For the Lam\'e system, we rescale the Lam\'e equation (\ref{lame}) as
\begin{eqnarray}
-\frac{1}{\kappa^2}\, \frac{d^2}{d\phi^2}\,  \psi+\nu \, {\rm sn}^2(\phi|\nu)\, \psi =
\left(\frac{E}{\kappa^2}+\frac{1}{2}\right)\, \psi\,,
\label{rescaledlame}
\end{eqnarray}
and, thus,  identify $\hbar$ in Eq. (\ref{disc}) as
\beq
\frac{1}{\kappa}\leftrightarrow \hbar\,.
\label{kappahbar}
\eeq
As a result, the condition for the occurrence of the
$J$-th gap takes the form
\beq
\kappa\, \, \sum_{n=0}^\infty \,
\frac{1}{\kappa^n }  \, {S_n\, (2K)} = J\,\pi= 
\pi\, \kappa\left(1-\frac{1}{2\,
\kappa}+\frac{1}{8\,\kappa^2}-\frac{1}{128\,\kappa^4} +
\dots \right)\,,
\label{jjj}
\eeq
where $2K$ is the period of the Lam\'e potential, and where on the right-hand side
we have expressed $J$ in terms of the effective semiclassical  expansion parameter
$1/\kappa$. This relation (\ref{jjj}) can be used to find an expansion for the
energy of the $J$-th gap by expanding
\begin{eqnarray}
E=\frac{1}{2}\,\kappa^2 +\sum_{\ell=1}^\infty\, 
\frac{\varepsilon_\ell}{\kappa^{\ell-2}}\,.
\label{eexp}
\end{eqnarray}
The expansion coefficients $\varepsilon_l$ are fixed by identifying terms on both
sides of the expansion in (\ref{jjj}). For example, the leading terms clearly match
because
\begin{eqnarray}
S_0\, (2K) &=& \int_0^{2K}\sqrt{\frac{E}{\kappa^2}+\frac{1}{2}-\nu\,{\rm
sn}^2(\phi|\nu)} \,\, d\phi
\nonumber\\[3mm]
&=& \int_0^{2K}\sqrt{1-\nu\, {\rm sn}^2(\phi|\nu)}\,\,
d\phi+O\left(\frac{1}{\kappa}\right)
\nonumber\\[2mm]
&=&\pi+O\left(\frac{1}{\kappa}\right)\,.
\label{leading}
\end{eqnarray}
The next-to-leading term on the left-hand-side of the expansion (\ref{jjj}) comes
from expanding $S_0(2K)$:
\begin{eqnarray}
S_0\, (2K) &=& \int_0^{2K}\sqrt{1-\nu\,{\rm
sn}^2(\phi|\nu)+\frac{\varepsilon_1}{\kappa }}\,\, d\phi+\dots
\nonumber\\[3mm]
&=&
\pi+\frac{\pi\,\varepsilon_1}{2\kappa\,
\sqrt{1-\nu}}+O\left(\frac{1}{\kappa^2}\right)\,.
\label{nexttoleading}
\end{eqnarray}
Matching the next-to-leading terms in (\ref{jjj}), we find
\begin{eqnarray}
\varepsilon_1=-\sqrt{1-\nu}\,.
\label{match}
\end{eqnarray}
This agrees with the next-to-leading term in the perturbative expansion
(\ref{e0pt}), after making the duality transformations $\nu\to 1- \nu$ and $E\to
-E$.

Note that there is no contribution from $S_1$, since $Q(x)$ is periodic with period
$2K$. (It is important here that the turning points are off the real axis, so that
integrating along the real period we do not encounter any turning points, which
would then require deformation of the integration contour in the complex plane
\cite{dunham,carlbook}). Indeed, for the same reason, none of the odd-indexed
$S_n$ contributes to the discriminant.

Similarly, successive orders in the energy expansion (\ref{eexp}) follow by
expanding (\ref{jjj}) to a given order in $1/\kappa$, and matching to determine the
expansion coefficients
$\varepsilon_\ell$. In this way one finds
\begin{eqnarray}
E=\frac{1}{2}\,\kappa^2
-\kappa\,
\sqrt{1-\nu}
+\frac{2-\nu}{4}+\frac{(-2+2\nu+\nu^2)}{16\,\kappa\,\sqrt{1-\nu}}
+\dots\,.
\label{wkbexp}
\end{eqnarray}
Comparing with the perturbative expansion (\ref{e0pt}) we see that the semiclassical expansion
(\ref{wkbexp}) is indeed the dual of the perturbative expansion (\ref{e0pt}), under
the duality transformation $\nu\to 1-\nu$ and $E\to -E$.

\subsection{Estimates of widths of bands and gaps}
\label{widths}

The semiclassical techniques used in the previous section were not
sensitive to the exponentially small corrections needed to estimate the
{\it width} of a low-lying band, or a high-lying gap. Yet, by the duality
transformation (\ref{duality}), the width of the lowest band is equal
to the width of the highest gap, for the dual potential obtained by making
the replacement $\nu\to 1-\nu$.  In this section we compute the widths of the
lowest band and highest gap, using various approximate techniques, and
compare with the exact duality result. This provides a link between
perturbative and non-perturbative techniques that is much more sensitive
than that discussed in the previous section for the {\it locations} of
low-lying bands and high-lying gaps.

\subsubsection{Algebraic approach for width of lowest band}
\label{algbandwidth}

Since the band edge energies are given by the eigenvalues of the finite
dimensional matrix $H$ in (\ref{ham}), the most direct way to evaluate the
width of the lowest-lying band is to take the difference of the two smallest
eigenvalues of $H$. For any given value of the elliptic parameter $\nu$,
this involves finding the eigenvalues of a $(2J+1)\times (2J+1)$ matrix,
which can be done with great precision. However, finding
the eigenvalues as {\it functions} of $\nu$ (see, for example, the plots
in Figs.~\ref{ppf3},
\ref{ppf4}, \ref{ppf1} and \ref{ppf2} for
$J=2$ and $J=5$) rapidly becomes complicated as
$J$ increases. Nevertheless, from these expressions it is possible to
deduce \cite{dr} the exact leading behavior, in the limit $\nu\to 1$, of
the width of the lowest band, for any $J$ :
\begin{eqnarray}
\Delta E^{\rm algebraic}_{\rm band} = {8J\, \Gamma(J+1/2)\over 4^J
\sqrt{\pi}\,
\Gamma(J)} \, (1-\nu)^J\, \left(1+\frac{J-1}{2} (1-\nu)+\dots\right)
\label{exact}
\end{eqnarray}
This clearly shows the exponentially narrow character of the lowest band in
the $\nu\to 1$ limit.

\subsubsection{Tight-binding approximation for width of lowest band}
\label{tbbandwidth}

In the tight-binding approximation \cite{peierls}, one assumes
the periodic wells are far separated, so that the periodic potential can be
treated as a periodic sequence of ``atomic" wells, each of which has a set
of discrete bound levels. Small overlap effects broaden these discrete
bound levels into bound bands. In the Lam\'e case this approximation can
be made very explicit due to the remarkable elliptic function identity:
\begin{eqnarray}
\nu\, {\rm sn}^2(\phi |\nu)
=\frac{E^\prime}{K^\prime}-
\left(\frac{\pi}{2K^\prime}\right)^2
\sum_{n=-\infty}^\infty 
\left[ {\rm sech}\left(\frac{\pi}{2K^\prime}(\phi-2 n K)\right)\right]^2\,.
\label{magic}
\end{eqnarray}
Here  $E(\nu)$ is the complete elliptic
integral of the second kind \cite{ww,abramowitz}, and
$$
E^\prime(\nu)=E(1-\nu)\,.
$$ 
This identity (\ref{magic}) shows that the
periodic Lam\'e potential can be written as a sequence of periodically
displaced (by the period $2K(\nu)$) P\"oschl-Teller ``atomic" wells (but
the rescaling factor
$\frac{\pi}{2 K^\prime}$ is non-obvious). This ``atomic" structure becomes
clear graphically as $\nu\to 1$ (see Fig.~\ref{ppf6}), but is in fact true for all
$\nu$. Since each ``atomic" well is a P\"oschl-Teller well, the normalized
lowest energy bound state is 
\begin{eqnarray}
\Psi_0(\phi)=\sqrt{\frac{\sqrt{\pi}\, \Gamma(J+1/2)}{2K^\prime\, \Gamma(J)}}  
\,\, \left[ {\rm sech} \left( \frac{\pi}{2K^\prime}\, \phi\right) \right]^J\,.
\label{ptground}
\end{eqnarray}
In the tight-binding (TB) approximation, the energy of this lowest state
broadens into a band of width \cite{dr,peierls} 
\begin{eqnarray}
\Delta E^{\rm TB}_{\rm band} \!\! &=&\!\!
 4 \kappa^2  \left( \frac{\pi}{2K^\prime}\right)^2
\int_{-\infty}^\infty d\phi \sum_{n\neq 0}
\left[{\rm sech}
\left(\frac{\pi (\phi-2 n K)}{2K^\prime}\right) \right]^2\, 
\Psi_0(\phi) \Psi_0(\phi-2 K)\nonumber\\[3mm] 
&\approx & {8J\, \Gamma(J+1/2)\over 4^J \sqrt{\pi}\,
\Gamma(J)} \,\, (1-\nu)^J\, \left(1+\frac{J-1}{2} (1-\nu)+\dots\right)\,,
\label{tb}
\end{eqnarray} 
where in the second line we have kept dominant terms as
$\nu\to 1$, and used the fact that 
$$
\exp\left(-\pi \frac{ K}{K^\prime}\right) \sim
\frac{1-\nu}{16} \left(1 +\frac{1}{2}(1-\nu)+\dots\right)\,.
$$
 This result (\ref{tb})
agrees precisely with the exact result (\ref{exact}) to this order in
$1-\nu$. Therefore, for any $J$, the tight-binding approximation is good
as $\nu\to 1$; {\it i.e.} as the separation between atomic wells becomes
large.

\subsubsection{Instanton approximation for width of lowest band}
\label{instbandwidth}

In the instanton approximation \cite{zinn}, tunneling is suppressed because
the barrier height is much greater than the ground state energy of any given
isolated ``atomic" well. For the Lam\'e potential in (\ref{lame}), this
means 
\begin{eqnarray}
\kappa\, \sqrt{\nu}\gg 1 \,.
\label{semi}
\end{eqnarray}
Remarkably, the instanton calculation for the Lam\'e potential can be done in
closed form \cite{dr}, leading to 
\begin{eqnarray}
\Delta  E^{\rm instanton}_{\rm band}=\frac{16}{\sqrt{\pi}} 
\, \left(\kappa^2\, \nu
\right)^{3/4} \, \left(1+\sqrt{\nu}\right)^{-2\kappa }\,
\left(1-\nu\right)^{\kappa- \frac{1}{2}} \,.
\label{instanton}
\end{eqnarray} 
To compare this instanton result (\ref{instanton}) with the algebraic result
(\ref{exact}) we take $\nu\to 1$, and we take $J$ to be large, in order to be
in the semiclassical limit (\ref{semi}). Then 
\begin{eqnarray}
\Delta  E^{\rm instanton}_{\rm band}\sim \frac{8 J^{3/2}}{\sqrt{\pi}\, 4^J}
\,\,
\left(1-\nu\right)^J \left[ 1+\frac{J-1}{2}(1-\nu)+\dots \right]\,,
\label{largej}
\end{eqnarray} 
which agrees perfectly with the large $J$ limit (using Stirling's formula) of
the exact algebraic result (\ref{exact}). Thus, this example gives an
analytic confirmation that the instanton calculation gives the correct
leading large $J$ behavior of the width of the lowest band, as $\nu\to 1$.
For other values of $\nu$, the instanton formula (\ref{instanton}) is also
the correct leading large $J$ result, but the comparison 
with $\Delta E^{\rm algebraic}_{\rm band}$ must be done
numerically \cite{dr} for a given value of $\nu$  since the large $J$ asymptotic
behavior of $\Delta E^{\rm algebraic}_{\rm band}$ for arbitrary $\nu$ is not
analytically calculable.

\subsubsection{WKB approximation for width of lowest band}
\label{wkbbandwidth}

We can also estimate the width of the lowest band using WKB.
First, rescale the the Lam\'e equation (\ref{lame}) as
\beq
 -\frac{1}{2}\, \frac{1}{\kappa^2\, \nu  }\, \frac{d^2 \Psi}{d\phi^2}
+\frac{1}{2}\, {\rm sn}^2\, (\phi|\nu )\, \Psi (\phi) = 
\left(\frac{1}{4\nu}+
\frac{E}{2\, \kappa^2\,\nu}\right)
\Psi (\phi)\equiv {\cal E}\, \Psi (\phi) \,.
\label{lamemod}
\eeq
 In this form we clearly see the small WKB
parameter to be 
$$``\hbar"=\frac{1}{\kappa\, \sqrt{\nu}}\,,$$ 
so that the large $J$ limit
is indeed semiclassical. Then, the standard WKB
\cite{landau,gildener,dr} expression for the band width is
\begin{eqnarray}
\Delta {\cal E}^{\rm WKB}=\frac{2``\hbar"}{\pi}
\,\exp\left(-\frac{1}{``\hbar"}\, \int_{\rm t.p.}\, d\phi\,  \sqrt{2(V(\phi)-{\cal
E})}\right)
\label{landauwkb}
\end{eqnarray}
where  t.p.  denotes the turning points. Thus, remembering to rescale the
energy and using the first two terms in Eq. (\ref{e0pt}) for the
energy  of the lowest band, we get
\begin{eqnarray}
\Delta E^{\rm WKB}_{\rm band}=\frac{4}{\pi} \,\kappa\, \sqrt{\nu }
\,\exp\left\{ - {\kappa\, \sqrt{\nu }}\, \int_{\rm t.p.}\, d\phi \, 
\sqrt{{\rm sn}^2(\phi |\nu )- \frac{1}{\kappa\,\sqrt\nu} }\, \right\}\,.
\label{landau}
\end{eqnarray}

\vspace{1mm}

\noindent
Defining $y={\rm sn}^2(\phi |\nu )$, we can now evaluate the integral in the
semiclassical limit where $``\hbar"=1/(\kappa\, \sqrt{\nu }) \ll 1$,

\vspace{1mm}

\begin{eqnarray}
&& \int_{``\hbar"}^1
\frac{\sqrt{1-``\hbar"/y}}{\sqrt{1-y}\sqrt{1-\nu\,  y}}\, dy =
\frac{1}{\sqrt{\nu}}\,
\ln \left(\frac{1+\sqrt{\nu}}{1-\sqrt{\nu}} \right)  
\nonumber\\[3mm]
&&   -``\hbar"\left( \frac{1}{2}\ln 
``\hbar"+\frac{1}{2}\ln \, (1-\nu)- 2 \ln 
2-\frac{1}{2} \right)
+O\left( (``\hbar")^2\right) \,. 
\label{hb}
\end{eqnarray}

\vspace{1mm}

\noindent
As a result, one finds \cite{dr} from (\ref{landau})
\begin{eqnarray}
\Delta E^{\rm WKB}_{\rm band} &=&\sqrt{\frac{e}{\pi}} \, \Delta E^{\rm
instanton}_{\rm band}\,. 
\label{factor}
\end{eqnarray}
This factor of $\sqrt{e/\pi}$ has been known for a long time
\cite{goldstein}. (The instanton result for $\Delta E_{\rm band}$
is correct while Eq. (\ref{landau}) is off by this factor.)
It has been found in many other comparisons
between the instanton method and the WKB formula (\ref{landauwkb})
\cite{gildener,neuberger,leonardis,regina}. It can be traced to a slightly
crude matching of normalizations of wave functions
\cite{goldstein,furry,dh}. A more precise WKB treatment leads, of course,
to complete agreement between these two semiclassical methods, as has
recently been emphasized explicitly for the Lam\'e potential \cite{muller}.

\subsubsection{Algebraic approach for width of highest gap}
\label{alggapwidth}

We now turn our attention to the width of the highest {\it gap}.
Taking the difference of the two largest eigenvalues of the finite
dimensional matrix $H$ in (\ref{ham}), it is straightforward to show that as
$\nu\to 0$, for any $J$, this difference gives 
\begin{eqnarray}
\Delta E^{\rm algebraic}_{\rm gap} = {8J\, \Gamma(J+1/2)\over 4^J
\sqrt{\pi}\,
\Gamma(J)} \, \nu^J \left(1+\frac{J-1}{2}\, \nu+\dots\right)\,,
\label{exacttop}
\end{eqnarray}
which is the same as the algebraic expression (\ref{exact}) for the width
of the lowest band, with the duality replacement $\nu\to 1-\nu$.

This result illustrates a theorem due to Trubowitz \cite{trubowitz}, which
states that for a real analytic periodic potential, the gap widths shrink
exponentially fast as one goes up in the spectrum. This in turn extends an
earlier result of Hochstadt \cite{hochstadt}, which states that the gap
widths go like $0(\frac{1}{l^m})$ if the potential is
$m$ times differentiable. For a real analytic potential, such as the Lam\'e
potential in (\ref{lame}), the gap widths shrink faster than any power of
the gap label $l$, and, in fact, shrink exponentially. By duality, the
low-lying bands are also exponentially narrow, as in (\ref{exact}).

\subsubsection{Naive perturbative estimate for width of highest gap}
\label{nptgapwidth}

As $\nu\to 0$, for fixed $J$, the potential in (\ref{lame})  becomes
weak. Thus, one should be able to solve this problem using perturbation
theory. Near $\nu=0$, the $n$-th  gap can be considered as a splitting
between the two degenerate free states $\Psi_\pm=\exp(\pm in\phi)$ at
$E=n^2-\frac{1}{2}J(J+1)$. A standard solid state physics approximation gives
\cite{peierls}
the width of the
$n$-th gap as (twice) the
$n$-th Fourier component of the potential,
\begin{eqnarray}
\Delta E^{n}_{\rm gap}\approx 2 |V_n|\,.
\label{peierls}
\end{eqnarray}
The Jacobi elliptic function has Fourier decomposition \cite{ww}
\begin{eqnarray}
\nu \, {\rm sn}^2(\phi|\nu)=1-\frac{E(\nu)}{K(\nu)}+\frac{\pi^2}{K^2}\, 
\sum_{n=1}^\infty \,\, \frac{n\,
\cos\left(\frac{n\pi\phi}{K}\right)}{\sinh\left(\frac{n\pi
K^\prime}{K}\right)}\, .
\label{fourier}
\end{eqnarray}
Thus, for the highest (i.e., the $J$-th) gap, we would deduce
\begin{eqnarray}
\Delta E^{J}&\approx& J(J+1)\,  \frac{\pi^2}{K^2}\, \, 
\frac{J}{\sinh\left(\frac{J\pi
K^\prime}{K}\right)}\nonumber\\[3mm]
&\sim & 8\, J^2(J+1) \, \left(\frac{\nu}{16}\right)^J\, ,\qquad \nu\to 0\,.
\label{wrong}
\end{eqnarray}
When $J=1$, this agrees with the leading term in the  exact algebraic result
(\ref{exacttop}). But it does not agree when $J\geq 2$. This failure
illustrates an important point -- the formula (\ref{peierls}) comes from
first-order in perturbation theory. However, from (\ref{exacttop}) we see
that the width of the highest gap is of $J$-th order in perturbation
theory. So first-order perturbation theory is clearly not sufficient for
$J\geq 2$. Indeed, to compare with the semi-classical (large
$J$) results for the width of the lowest band, we see that we will have to be
able to go to very high orders in perturbation theory. This is an
interesting, and very direct, illustration of the well-known connection
between non-perturbative physics and high orders of perturbation theory.

\subsubsection{Perturbation theory to order $J$ for width of highest gap}
\label{ptgapwidth}

It is generally very difficult to go to high orders in perturbation theory,
even in quantum mechanics. For the Lam\'e system (\ref{lame}) we can
exploit the algebraic relation to the finite-dimensional spectral problem
(\ref{ham}). However, since $H$ in (\ref{ham}) is a $(2J+1)\times (2J+1)$
matrix, the large $J$ limit is still non-trivial. Since we are only
interested in gap widths, in this section we neglect the constant term in
$H$, and treat $\nu J_y^2$ as a perturbation of the free matrix $J_x^2$,
which has eigenvalues $0, 1^2, 2^2,\dots, J^2$. The non-zero eigenvalues
of $J_x^2$ are two-fold degenerate, while the zero eigenvalue is
nondegenerate. We are interested in the splitting of the two degenerate
eigenvectors of $J_x^2$ with the highest eigenvalue, $J^2$. Even though
these two states are degenerate, they do not mix in perturbation theory,
because of the form of the perturbation matrix $J_y^2$. This is
essentially because both
$J_x^2$ and
$J_y^2$ have a sub-block structure in which alternate rows and columns
separate. For definiteness, we choose the following
$(2J+1)\times (2J+1)$ representation:
\begin{eqnarray}
J_x&=&\frac{1}{2}\left(\matrix{0&\sqrt{2J}&0 \cr \sqrt{2J}&0&\sqrt{2(2J-1)}&0 \cr
0&\sqrt{2(2J-1)}&0&\sqrt{3(2J-2)} \cr
&0&\ddots&\ddots&\ddots \cr
&&&&&\sqrt{2J}\cr
&&&0&\sqrt{2J}&0}\right) ,\nonumber\\[6mm]
J_y&=&-\frac{i}{2}\left(\matrix{0&\sqrt{2J}&0 \cr 
-\sqrt{2J}&0&\sqrt{2(2J-1)}&0 \cr
0&-\sqrt{2(2J-1)}&0&\sqrt{3(2J-2)} \cr
&0&\ddots&\ddots&\ddots \cr
&&&&&\sqrt{2J}\cr
&&&0&-\sqrt{2J}&0}\right) ,\nonumber\\[6mm]
J_z&=&{\rm diag}\left(J,J-1,J-2, \dots, -J+2,-J+1,-J\right)\,. 
\label{rep}
\end{eqnarray}
Then the matrix elements of
$J_y^2$ between the orthonormal eigenvectors of $J_x^2$ have the simple
sparse structure

\begin{eqnarray}
\tiny{\langle J_y^2\rangle=\left(
\matrix{L_1&0 &0& O_0&0 &0&\dots&&&&&&&&0 \cr
0&L_2&0&0&0&O_1&0&&&&&&&&\cr
0&0&L_3&0&0&0&O_1&0\cr
O_0&0&0&D_1&0&0&0&O_2&0\cr
0&0&0&0&D_1&0&0&0&O_2&0\cr
0&O_1&0&0&0&D_2&0&0&0&\dots&0 \cr
&0&O_1&0&0&0&D_2&0&0&0&\ddots&0 \cr
&&0&O_2&0&0&0&\ddots &0&0&0&O_{J-3}&0&&\cr
&&&0&O_2&0&0&0&\ddots &0&0&0&O_{J-3}&0& \cr
&&&&0&\ddots&0&0&0&D_{J-3}&0&0&0&O_{J-2}&0\cr
&&&&&0&\dots&0&0&0&D_{J-3}&0&0&0&O_{J-2}  \cr
&&&&&&0&O_{J-3}&0&0&0&D_{J-2}&0&0&0\cr
&&&&&&&0&O_{J-3}&0&0&0&D_{J-2}&0&0\cr
&&&&&&&&0&O_{J-2}&0&0&0&D_{J-1}&0\cr
0&&&&&&&&&0&O_{J-2}&0&0&0&D_{J-1}}\right)} 
\nonumber\\
\label{matrixelts}
\end{eqnarray}
where the nonzero entries in the top left corner are 
\begin{eqnarray}
L_1 &=& \frac{1}{2}J(J+1)\,,\nonumber\\[1mm]
L_2 &=& \frac{1}{4}J(J+1)-\frac{1}{2}\,,\nonumber\\[1mm]
L_3 &=& \frac{3}{4}J(J+1)-\frac{1}{2}\,.
\label{topleft}
\end{eqnarray}
The remaining diagonal entries appear in $2\times 2$ diagonal blocks, ${\rm
diag}(D_n,D_n)$, with
\begin{eqnarray}
D_n=\frac{1}{2}\, J(J+1)-\frac{1}{2}\, (n+1)^2 \,, \qquad n=1,2, ..., J-1\,.
\label{diags}
\end{eqnarray}
Finally, the nonzero off-diagonal entries in (\ref{matrixelts}) are 
\begin{eqnarray}
O_0&=&-\frac{1}{4}\sqrt{2J(J+2)(J^2-1)}\,, \nonumber\\[3mm]
O_n&=&-\frac{1}{4}\sqrt{(J-n)(J+n+2)(J^2-(n+1)^2)}\,,\quad n=1,\dots, J-2\,.
\label{offdiag}
\end{eqnarray}

\vspace{1mm}

\noindent
In the basis used for (\ref{matrixelts}), the last two rows and columns refer
to the degenerate states with the highest eigenvalue of $J_x^2$. We see that
there is indeed no string of matrix elements connecting $\langle
J_y^2\rangle_{2J+1,2J+1}$ with
$\langle J_y^2\rangle_{2J,2J}$. Thus, these two degenerate states do not mix
at any order of perturbation theory.

\vspace{1mm}

Consider perturbing around the $(2J)$-th eigenstate of $J_x^2$, with
eigenvalue $J^2$. The energy $E_{2J}$ of the perturbed system can be
expressed implicitly as \cite{mathews}
\begin{eqnarray}
E_{2J}=E^0_{2J}+U_{2J,2J}+\sum_{n\neq
2J}\frac{U_{2J,n}U_{n,2J}}{E_{2J}-E^0_{n}}+\sum_{n\neq
2J}\sum_{m\neq
2J}\frac{U_{2J,n}U_{n,m}U_{m,2J}}{(E_{2J}-E^0_{n})(E_{2J}-E^0_{m})}+\dots
\nonumber\\
\label{allorders}
\end{eqnarray}
where $E^0_{n}$ is the $n$-th eigenvalue of the free system, and
$U_{m,n}$ denotes the matrix element of the perturbation between the $m$-th and
$n$-th free states. An analogous formula applies for
$E_{2J+1}$. 

\vspace{1mm}

>From the $2\times 2$ sub-block structure of (\ref{matrixelts}), it is
clear that for the terms in (\ref{allorders}) with strings of fewer than
$J$ matrix element factors, the shifts in $E_{2J}$ and $E_{2J+1}$ are
identical. Thus, these terms are irrelevant for the computation of the
{\it difference} between $E_{2J}$ and
$E_{2J+1}$. So, we can concentrate solely on the $J$-th order term in
(\ref{allorders}), which involves a product of exactly $J$ matrix element
factors. Also, since we are looking for the leading term at this order of
perturbation theory, we can replace the true eigenvalue $E_{2J}$ or
$E_{2J+1}$ in the denominator by the free eigenvalue, which is $J^2$. A further
simplification follows from the fact that there is only {\bf one} nonzero
string of $J$ matrix element factors beginning and ending with the index
$2J$ or $2J+1$. Thus, to compute the difference $E_{2J+1}-E_{2J}$ at
$J$-th order, we simply have to take the difference between the
string of matrix elements in each case, with the appropriate denominator
factor.

To proceed we need to specify whether $J$ is even or odd. We consider
$J$ to be even (the case of $J$ odd only requires simple
modifications, and is left to the reader). With $J$ even, there is in fact no
nonzero string of $J$ matrix elements beginning and ending at $2J+1$.
However, there is such a string beginning and ending at $2J$,
\begin{eqnarray}
&& \left[ U_{2J,2J-4}U_{2J-4,2J-8}\dots
U_{8,4}U_{4,1}\right]^2 = 2\prod_{l=1,3,\dots,
J-1}\left[\left(\frac{1}{4}\right)^2l(l+1)(2J+1-l)(2J-l)\right]
\nonumber\\[3mm]
&&=\frac{2}{4^J}\left[ J! (2J)(2J-1)(2J-2) \dots (J+2)(J+1)\right] 
= \frac{2(2J)!}{4^J}\,. 
\label{numerator}
\end{eqnarray}
This, together with a factor of $\nu^J$, gives the numerator of the
$J$-th order term in the perturbation series (\ref{allorders}). The
denominator of  the $J$-th order term, with $E_{2J}$ replaced by
$J^2$, is
\begin{eqnarray}
\left[(J^2-(J-2)^2)(J^2-(J-4)^2)\dots (J^2-4)\right]^2 J^2 = 4^{J-1}
\left[(J-1)!\right]^2\,.
\label{denominator}
\end{eqnarray}
Therefore, the splitting between $E_{2J+1}$ and $E_{2J}$ is given, at 
order $J$ in perturbation theory, by the ratio of (\ref{numerator}) to
(\ref{denominator}),
\begin{eqnarray}
\Delta E^{\rm pert. theory}_{\rm gap}=\frac{8}{4^{2J}} \,\,
\frac{(2J)!}{\left[(J-1)!\right]^2}\,\nu^J= \frac{8 J\, \Gamma(J+1/2)}{4^J \sqrt{\pi}\,
\Gamma(J)}\, \nu^J\,,
\label{pt}
\end{eqnarray}
in complete agreement with the leading part of the exact algebraic result
(\ref{exacttop}). This gives the leading $\nu$ dependence of the width of
the highest gap, for any $J$. This provides an explicit illustration of
the connection between the non-perturbative results of the previous
section, and high orders of perturbation theory.

\subsubsection{WKB and ``over the barrier tunneling" for width of highest
gap}
\label{wkbgapwidth}

As mentioned in section \ref{wkbgaplocation}, in the semiclassical limit,
the highest gap lies {\it above} the top of the potential, so we need to use
``over-the-barrier" WKB to determine the width of the highest gap. This
means that the turning points lie off the real axis, and the  width of the
gap is given by the expression
\cite{landau}
\begin{eqnarray}
\Delta E^{\rm WKB}_{\rm gap} =\frac{2 }{\pi}\,\exp\left(- 2\, {\rm  Im}\,
\int_{C} p\, dx\right)\,,
\label{wkbover}
\end{eqnarray}
where $p(\phi)=\sqrt{2(E-V(\phi))}$, and the integration is over a contour
${\cal C}$ beginning on the real $\phi$ axis, then encircling one of the
complex turning points, and returning to the real axis. The complex
turning points lie on the vertical line through the point $\phi=K$.
Thus, it is convenient to change variables to 
\begin{eqnarray}
\phi=K+iK^\prime-i\phi^\prime\,.
\label{vertical}
\end{eqnarray}
This is precisely the duality transformation change of variables
(\ref{rotation}) discussed in Sect.~\ref{introduality}. Thus
\cite{ww,abramowitz},
\begin{eqnarray}
E-V(\phi|\nu)&\equiv &E+\frac{1}{2}J(J+1)-J(J+1)\,\nu\, {\rm sn}^2(\phi|\nu)
\nonumber\\[1mm]
&=& E -\frac{1}{2}J(J+1)+J(J+1)\,(1-\nu)\, {\rm sn}^2(\phi^\prime|1-\nu)
\nonumber\\[2mm]
&=&V(\phi^\prime|1-\nu)-(-E)\,.
\label{wkbdual}
\end{eqnarray}
Therefore, since the location of the highest gap is approximately (see
(\ref{wkbexp}))
$$
E=\frac{1}{2}J(J+1)-\sqrt{(1-\nu)\, J(J+1)}\,,
$$ 
we see that the evaluation of
the WKB integral in the exponent of (\ref{wkbover}) is {\it identical} to the
WKB integral (\ref{hb}) computed in Sect. \ref{wkbbandwidth} for the
width of the lowest band, except for the replacement $\nu\to 1-\nu$. Thus,
the WKB estimate for the width of the highest gap is identical to the WKB
estimate for the width of the lowest band, with the duality transformation
$\nu\to 1-\nu$.

\section{Conclusions}
\label{conclu}

In summary, we have shown that for the QES periodic Lam\'e system  (\ref{lame}),
there is a duality symmetry that maps the spectrum into the 
energy-reflected spectrum of the dual Lam\'e potential which has dual elliptic 
parameter $\nu^\prime=1-\nu$. This means that bands or gaps high (low) in the 
spectrum are mapped into gaps or bands low (high) in the dual spectrum. The 
self-dual point, $\nu=\frac{1}{2}$, of this duality transformation corresponds 
to an  energy reflection symmetry of the self-dual potential. This also 
provides an extension of the energy-reflection construction of \cite{st2}.
Furthermore, this approach decomposes the $(2J+1)\times (2J+1)$ 
algebraic spectral problem into four smaller algebraic problems (the 
precise details of this
decomposition depend on whether $J$ is odd or even). This 
decomposition represents a significant algebraic simplification in the large $J$ 
limit.  The large
$J$ limit is a semiclassical limit, and we have also shown in detail 
how the duality transformation relates the weak coupling
(perturbative) and semiclassical (nonperturbative) 
sectors. Such a relation arises because the energy reflection aspect of the 
duality  symmetry relates states high up in the spectrum to states low 
down in the spectrum. Interestingly, since the QES models discussed 
here are  periodic, and hence have band spectra, this duality applies not 
just to the locations of the bands and gaps, but also to the widths of the 
bands and gaps.  The calculations of such widths are sensitive to exponentially 
small contributions that are neglected in the previous calculations of 
locations of energy levels \cite{bdm,st2,kavic}. We were able to show that 
  duality  applies also to the calculation of 
these  exponentially small effects.

There are several directions in which it would be interesting to generalize
this analysis. It would  be interesting to find such a duality in 
models in more than one 
dimension. There is presumably a more
geometrical interpretation of this duality, in terms of the associated
genus $J$ Riemann surface, and this may provide a useful alternative
viewpoint which extends to other finite gap potentials. Finally, the Lam\'e
models possess other interesting transformation properties which remain to
be explored from the QES perspective. For example, since
\begin{eqnarray}
J_x^2+\nu J_y^2=\nu J(J+1)\,{\bf I}+(1-\nu)\left(J_x^2-\frac{\nu}{1-\nu}
J_z^2\right)
\label{inv}
\end{eqnarray} 
the band edge energies for elliptic parameter $\nu$ are related to those
for elliptic parameter $-\frac{\nu}{1-\nu}$, which reflects the behavior
of the Jacobi function under such a parameter change. It would be
interesting to investigate these more general modular transformations in the
context of quasi-exact solvability.

\section*{Acknowledgments}

We thank C. Bender, A. Turbiner, A. Vainshtein and  M. Voloshin for valuable
discussions, and T. ter Veldhuis and M. Voloshin for assistance with
Mathematica. G.D. is supported in part by the DOE grant
DE-FG02-92ER40716, and M.S. is supported in part by the DOE grant 
DE-FG02-94ER408.

\end{document}